\documentclass[aps,a4paper,preprint,showkeys]{revtex4}%
\usepackage{amsfonts}
\usepackage{amsmath}
\usepackage{amssymb}
\usepackage{graphicx}%
\providecommand{\U}[1]{\protect\rule{.1in}{.1in}}
\begin{document}
\title{Viscous dynamics of drops and bubbles in Hele-Shaw cells: drainage, drag
friction, coalescence, and bursting}
\author{Ko Okumura (okumura.ko@ocha.ac.jp)}
\affiliation{Physics Department, Faculty of Science, Ochanomizu University}

\begin{abstract}
In this review article, we discuss recent studies on drops and bubbles in
Hele-Shaw cells, focusing on how scaling laws exhibit crossovers from the
three-dimensional counterparts and focusing on topics in which viscosity plays
an important role. By virtue of progresses in analytical theory and high-speed
imaging, dynamics of drops and bubbles have actively been studied with the aid
of scaling arguments. However, compared with three dimensional problems,
studies on the corresponding problems in Hele-Shaw cells are still limited.
This review demonstrates that the effect of confinement in the Hele-Shaw cell
introduces new physics allowing different scaling regimes to appear. For this
purpose, we discuss various examples that are potentially important for
industrial applications handling drops and bubbles in confined spaces by
showing agreement between experiments and scaling theories. As a result, this
review provides a collection of problems in hydrodynamics that may be
analytically solved or that may be worth studying numerically in the near future.

\end{abstract}
\maketitle
\tableofcontents

\section{Introduction}

When a bubble is injected in a glass of milk, the bubble rises up in the
liquid and comes up at the liquid-air interface making a hemispherical air
bubble encapsulated by a film of the liquid. The bubble stays at the interface
with keeping the hemispherical shape but with decreasing the thickness of the
liquid film due to drainage. After some waiting time, a hole is created and
grows in the thin film, which leads to disappearance of the bubble. Similar
processes are observed with a highly viscous liquid even with lavas during
volcanic eruptions and even in the absence of surfactants.

These simple phenomena familiar to everyone involve several elementary
processes, such as creation of a bubble, rising of a bubble, film drainage,
nucleation of a hole, and bursting of a thin film. We will explore these
phenomena and how dynamics change when the bubble is confined in between two
plates of a Hele-Shaw cell. Throughout this review, we limit ourselves to the
case in which there are no effects of surfactants. We also discuss the case in
which the air bubble is replaced with a fluid drop and the case of coalescence
of a drop with a bath of the same liquid in a Hele-Shaw cell.

For the readers' convenience, we here introduce some notations used in this
review although they are re-introduced in the main text below as we proceed:

\begin{itemize}
\item $\eta_{1}$ and $\rho_{1}$: viscosity and density of a fluid drop (a
bubble or a liquid drop).

\item $\eta_{2}$ and $\eta_{2}$: viscosity and density of another fluid
(liquid or air) surrounding the fluid drop.

\item $\Delta\rho=\left\vert \rho_{1}-\rho_{2}\right\vert >0$ the density
difference of the two fluids; in discussing a bubble this is practically equal
to the density of the surrounding fluid $\rho_{2}$.

\item $\gamma_{12}$ interfacial energy between the fluid drop and the
surrounding fluid; in discussing a bubble, this is equal to the surface
tension of the surrounding fluid $\gamma_{2}$ (in discussing a liquid drop in
air in Sec.VI, this is equal to the surface tension of the liquid drop
$\gamma_{1}$).

\item $\kappa_{12}^{-1}=\sqrt{\gamma_{12}/(\Delta\rho g)}$ capillary length
for the interface between the fluid drop and the surrounding liquid; in
discussing a bubble, this is practically equal to the capillary length for the
surrounding fluid $\kappa_{2}^{-1}=\sqrt{\gamma_{2}/(\rho_{2}g)}$
\end{itemize}

\section{Life time of a bubble}

\begin{figure}[ptb]
\includegraphics[width=0.45\textwidth]{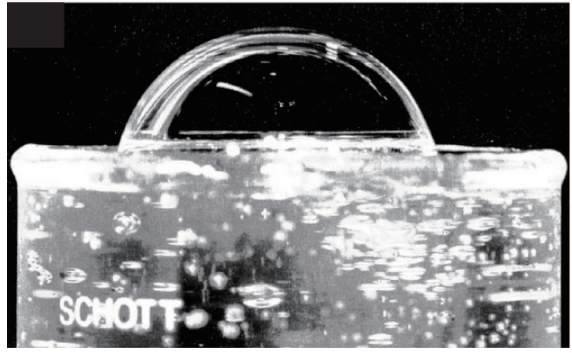}\caption{Hemispherical air
bubble formed at a liquid-air interface in a ultra-viscous liquid. Taken from
\cite{DebregeasGennesBrochard-Wyart1998}.}%
\label{f1}%
\end{figure}

When a bubble is created in a ultra-viscous liquid (viscosity $\eta_{2}$ and
density $\rho_{2}$), it raises up to make a hemispherical bubble at the
liquid-air interface as shown in Fig. \ref{f1}, and stays there for a
considerably long time, even though there are no effects of surfactants,
before it breaks up by bursting of the spherical thin film encapsulating the
bubble. In such a case, the life time of the bubble is governed by the
drainage of liquid from the thin film. With taking the $\theta$ coordinate as
the angle from the bubble top along the bubble surface ($\theta=0$ at the top
and $\theta\simeq\pi/2$ at the liquid-air interface), the thinning dynamics of
the film of thickness $h$ is known \cite{DebregeasGennesBrochard-Wyart1998} to
be given by%
\begin{equation}
h=h_{0}f(\theta)e^{-t/\tau} \label{e1}%
\end{equation}
for the initial thickness $h_{0}$ at the top of the bubble, i.e., at
$\theta=0$, with the life time $\tau$ and the function $f$ given by
\begin{align}
\tau &  =\eta_{2}/(\rho_{2}gR)\label{e2}\\
f(\theta)  &  =1/\cos^{4}(\theta/2) \label{e2b}%
\end{align}
with the radius of the bubble $R$ and the gravitational acceleration $g$.

This scaling law for the life time can be understood as follows. In the
present case, the velocity $\mathbf{v}$ satisfies the following lubricant
equation:%
\begin{equation}
\eta_{2}\nabla^{2}\mathbf{v}=-\rho_{2}\mathbf{g}, \label{e3}%
\end{equation}
where $\mathbf{v}=v(\theta)\mathbf{e}_{\theta}$ with $\mathbf{e}_{\theta}$ the
unit vector in the $\theta$ direction and the gravitational acceleration
vector $\mathbf{g}$ pointing in the direction $\theta=\pi$. This equation can
be solved in the form $v(\theta)\simeq(\rho_{2}gR^{2}/\eta_{2})\sin\theta$,
which results from the fact that the flow is a plug flow and does not change
in the direction of the film thickness (i.e., the direction of the $r$
coordinate for which $r=R$ at the bubble surface). The conservation equation,%
\begin{equation}
\frac{\partial h}{\partial t}+\nabla\cdot(h\mathbf{v)}=0, \label{e4}%
\end{equation}
can be solved through a separation of variables, which gives Eq. (\ref{e1})
with Eqs. (\ref{e2}) and (\ref{e2b}), by noting that $h=h(t,\theta)$ and
$\nabla\cdot(h\mathbf{v)=}\partial(\sin\theta hv)/\partial\theta/(R\sin
\theta)$.

\begin{figure}[ptb]
\includegraphics[width=0.8\textwidth]{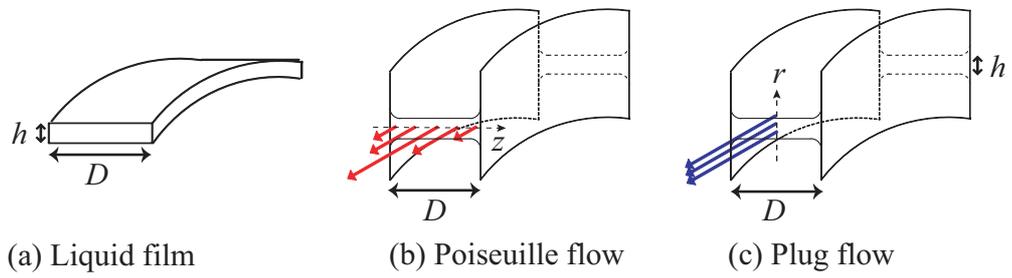}\caption{(a) A part of the
liquid film sandwiched by two cell plates separated by the distance $D$. In
the direction of cell thickness (i.e., the $z$ direction), the flow changes as
in (b). In the direction of film thickness (i.e., in the $r$ direction), the
flow does not change as in (c).}%
\label{f2}%
\end{figure}

When a bubble is sandwiched by two cell plates separated by the distance $D$
in a Hele-Shaw cell, the boundary condition changes as illustrated in Fig.
\ref{f2}. Because of this, Eq. (\ref{e1}) gives $v(\theta)\simeq(\rho
_{2}gD^{2}/\eta_{2})\sin\theta$ for $D<R$. In addition, the divergence
appearing in the conservation equation is changed into $\nabla\cdot
(h\mathbf{v)=}\partial(hv)/\partial\theta/R$. As a result, Eqs. (\ref{e2}) and
(\ref{e2b}) are replaced by
\begin{align}
\tau &  =12\eta_{2}R/(\rho_{2}gD^{2})\label{e5}\\
f(\theta)  &  =1/\cos^{2}(\theta/2) \label{e5b}%
\end{align}
for $D<R$. This theory is confirmed well in Ref. \cite{EriOkumura2007}.

\section{Drag friction acting on a bubble}

When a bubble is created in a liquid bath (viscosity $\eta_{2}$ and density
$\rho_{2}$), the bubble rises up in the liquid. When the rising velocity $V$
is small, we can assume that the shape of the rising bubble is practically a
spherical shape of radius $R$. In such a case, the drag friction acting on the
bubble is given by%
\begin{equation}
F\simeq\eta_{2}VR,\label{e6}%
\end{equation}
Note that Stokes' friction law for a solid sphere with radius $R$ moving with
a velocity $V$ in a viscous liquid with viscosity $\eta$ is given by
$F=6\pi\eta VR$.

\begin{figure}[ptb]
\includegraphics[width=0.9\textwidth]{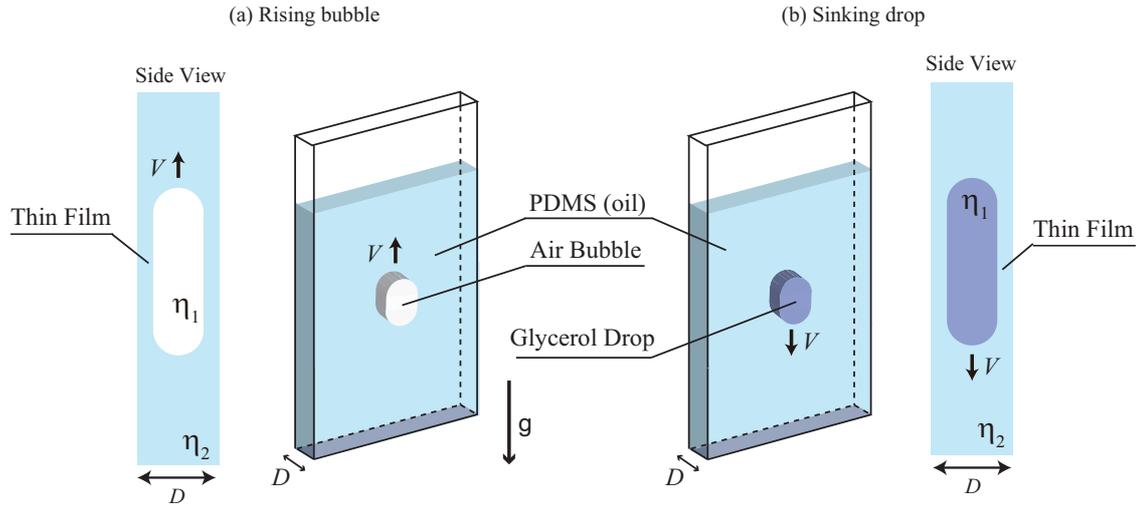}\caption{Illustrations of
experiments for a rising bubble (a) and for a sinking drop (b). The Hele-Shaw
cell is made up of transparent acrylic plates. The cell thickness $D$ is a few
millimeters and the size of the bubble or drop is about one centimeter. The
width and height of the cell are much larger than the size of the drop. Side
views show that the existence of liquid thin films between the fluid drop (the
bubble or liquid drop) and the cell walls. $\eta_{1}$ and $\eta_{2}$
respectively denote viscosities of the fluid drop and the liquid surrounding
the fluid drop.}%
\label{f3}%
\end{figure}

When a bubble is sandwiched by the two cell walls separated by the distance
$D$ of a Hele-Shaw cell, the drag force is found \cite{EriSoftMat2011} to be
given by%
\begin{equation}
F\simeq\eta_{2}VR_{T}^{2}/D \label{e7}%
\end{equation}
The experimental set up to confirm this is shown in Fig.~\ref{f3}(a), together
with a related set up discussed later. The air drop is slightly elongated,
which is characterized by the transverse and longitudinal lengths $R_{T}$ and
$R_{L}$ as defined in Fig.~\ref{f4}(a).

The scaling law given in Eq. (\ref{e7}) is explained by considering the slight
deformation of the bubble shape. When viscous dissipation is dominant for the
velocity gradient $V/D$ developed near the bubble between the cell plates
separated by $D$, the balance between the viscous dissipation and
gravitational energy gain (per unit time) can be expressed as
\begin{equation}
\eta_{2}(V/D)^{2}R_{T}^{2}D\simeq\rho_{2}gR_{T}R_{L}DV. \label{e8a}%
\end{equation}
This results in
\begin{equation}
V\simeq(R_{L}/R_{T})\rho_{2}gD^{2}/\eta_{2}, \label{e8}%
\end{equation}
which is discussed in \cite{TAYLORSAFFMAN1959} and \cite{Tanveer1986} as
further explained below. If Eq.~(\ref{e8a}) is expressed in the force-balance
form $\eta_{2}VR_{T}^{2}/D\simeq\rho_{2}gR_{T}R_{L}D$, we immediately know
that the drag friction acting on the bubble from the viscous medium is given
by Eq.~(\ref{e7}).

In order to observe this scaling law, dissipation associated with the velocity
gradient $V/D$ wins over other viscous dissipations, such as those associated
with velocity gradient $V/R_{T}$, and $V/R_{L}$, from which the condition
$D<R_{T}$, $R_{L}$ is required. An important fact to be noted here is that in
the present experiment, since the liquid surrounding the bubble completely wet
the cell plates, there are thin layers of the liquid between the cell plates
and both surfaces of the disk-shaped bubble (see the left side view in Fig.
\ref{f3}). The creation of velocity gradient $\simeq V/h$ with $h$ the
thickness of layers is energetically unfavorable due to the relation $h\ll D,$
and because of this the dissipation associated with the gradient $V/D$ becomes
dominant in the present case. However, this is not always the case, as shown
in Sec. \ref{s5}.

The scaling law in Eq. (\ref{e7}) is confirmed in Fig. \ref{f5}. In (a), the
rising velocity $V$ is given as a function of the cell thickness $D$ for
different parameters, $R_{T}$, $R_{L}$, and $\nu_{2}=\eta_{2}/\rho_{2}$. In
(b), the same data are plotted with renormalized axes, on the basis of the
dimensionless form of Eq. (\ref{e7}), $(R_{T}/R_{L})Ca=D/\kappa_{2}^{-1}$,
with the capillary number $Ca=\eta_{2}V/\gamma_{2}$ and the capillary length
$\kappa_{2}^{-1}=(\gamma_{2}/(\rho_{2}g))^{1/2}$, to show a clear data
collapse, confirming the validity of Eq. (\ref{e7}). Although the values of
the factor $R_{T}/R_{L}$ for the data used in Fig. \ref{f5} are relatively
close to one, this factor is important for the clear data collapse; it is
confirmed in \cite{EriSoftMat2011} that if this factor is set to one the
quality of the collapse is clearly deteriorated.

The rising bubble in a Hele-Shaw cell discussed above was theoretically
studied by Taylor and Saffman \cite{TAYLORSAFFMAN1959} (earlier than the
Bretherton's paper on bubbles in tubes \cite{Bretherton,Clanet2004}) and by
Tanveer \cite{Tanveer1986}. Other theoretical works on fluid drops in the
Hele-Shaw cell geometry includes the topological transition associated with
droplet breakup
\cite{Eggers1997,ConstantinDupontGoldsteinKadanoffShelleyZhou1993,GoldsteinPesciShelley1995,Howell1999}%
. \ 

Experimentally, a number of researchers have investigated the rising motion of
a bubble in a Hele-Shaw cell
\cite{Maxworthy1986,Kopf-SillHomsy1988,MaruvadaPark1996,HeleShawPetroleum2010}%
, together with studies performed in different geometries, which includes
studies on inertial regimes (Crossovers between viscous regimes treated in
this review and inertial and other regimes are fruitful future challenges)
\cite{PascalEPL2002,reyssat2014drops,CapsEPL2016}. However, they have mostly
concerned with the case in which there is a forced flow in the outer fluid
phase and/or the case in which the cell is strongly inclined nearly to a
horizontal position.\ 

One of the features of the present case is that thin liquid films surrounding
a bubble plays a crucial role as mentioned above (see the left side view in
Fig. \ref{f3}): in many previous works, the existence of such thin films is
not considered. In this respect, the present problem is closely related to the
dynamics governed by thin film dissipation such as the imbibition of textured
surfaces
\cite{StoneNM2007,IshinoReyssatEPL2007,ObaraPRER2012,TaniPlosOne2014,Tani2015,DominicVellaImbibition2016}%
, as further discussed in Sec. \ref{s5}. In this sense, our problem is quasi
two-dimensional, although the geometry of the Hele-Shaw cell is often
associated with a purely two-dimensional problem.

\section{Drag friction acting on a liquid drop}

Even if an air bubble is replaced by a liquid drop (density $\rho_{1}$) whose
viscosity $\eta_{1}$ is smaller than that of the surrounding bath liquid
$\eta_{2}$, the drag friction and the corresponding velocity are still given
respectively by Eqs. (\ref{e7}) and (\ref{e8}): $F\simeq\eta_{2}VR_{T}^{2}/D$
and $V\simeq(R_{L}/R_{T})\Delta\rho gD^{2}/\eta_{2}$. Note that, in the second
equation, $\rho_{2}$ is replaced with $\Delta\rho=\left\vert \rho_{2}-\rho
_{1}\right\vert $, which is practically equal to $\rho_{2}$ in the previous
section with a bubble of density $\rho_{1}$.

\begin{figure}[ptb]
\includegraphics[width=0.6\textwidth]{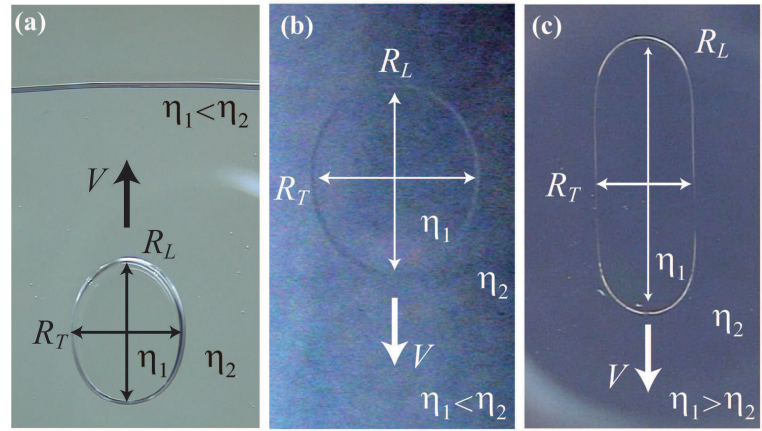}\caption{Moving fluid drops in
a Hele-Shaw cell. Here, the viscosity of the fluid drop (bubble or liquid
drop) and the surrounding fluid are denoted $\eta_{1}$ and $\eta_{2}$,
respectively. (a) The rising bubble ("air drop") is slightly deformed from a
circle and characterized by the longitudinal and transverse diameters, $R_{L}$
and $R_{T}$, respectively, as indicated in the photograph. (b) A glycerol drop
surrounded by a more viscous oil [as in (a)] is sinking in the cell. The fluid
drop shape is very similar to that in (a). (c) A glycerol drop surrounded by a
less viscous oil [different from (a) and (b)] is sinking in the cell. Compared
with (a) and (b), the elongation is significant. Figures are adapted from
\cite{EriSoftMat2011}.}%
\label{f4}%
\end{figure}

\begin{figure}[ptb]
\includegraphics[width=0.45\textwidth]{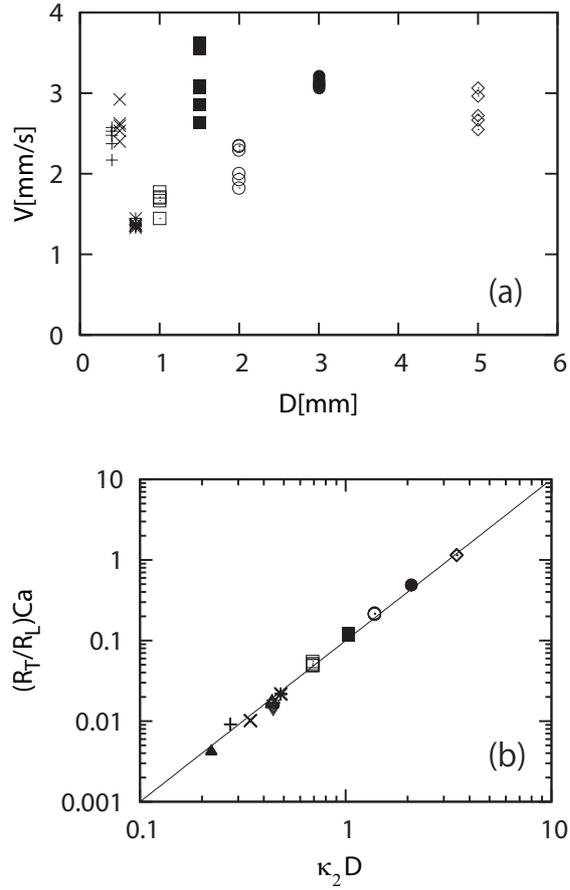}\caption{(a) $V$ vs $D$ for
different $R_{T},R_{L},$ and $\nu_{2}=\eta_{2}/\rho_{2}$. The 8 symbols +,
$\times,$+{\hspace{-3mm}}$\times,{\protect\small \square}%
,{\protect\small \blacksquare},{\protect\large \circ},{\protect\large \bullet
},$ and ${\protect\large \diamond}$ correspond to $(D$ [mm]$,\nu_{2}$
[cS]$)=(0.4,100)$, $(0.5,100)$, $(0.7,500)$, $(1.0,1000)$, $(1.5,1000)$,
$(2.0,3000)$, $(3.0,5000)$, and $(5.0,10000)$, respectively. (b) Plot in (a)
replotted with the two axes, the renormalized velocity $(R_{T}/R_{L})$Ca and
the renormalized cell thickness $\kappa_{2}D$, demonstrating a clear data
collapse by virtue of the theory. Figures are reproduced from
\cite{EriSoftMat2011}.}%
\label{f5}%
\end{figure}

The experimental setup for confirming this law is illustrated in
Fig.~\ref{f3}(b). In this case, the fluid drop goes down, because the density
of the drop is larger than that of the surrounding drop, and the drop shape is
considerably similar to that of an air bubble, as shown in Fig.~\ref{f4}(b).

However, when the drop viscosity $\eta_{1}$ is larger than that of the
surrounding liquid $\eta_{2}$ as in Fig.~\ref{f4}(c), the drag friction is
found to be given by the following scaling law \cite{EriSoftMat2011}:
$F\simeq\eta_{1}VR_{T}R_{L}/D$. This scaling law is expected because the
Eq.~(\ref{e8a}) is replaced by $\eta_{1}(V/D)^{2}R_{T}R_{L}D\simeq\Delta\rho
gR_{T}R_{L}DV$. In this case, the shape of the drop is significantly elongated
compared with the previous case.

In summary, two scaling regimes of the viscous drag friction acting on a fluid
drop in a Hele-Shaw cell ($R_{L},R_{T}>D$) are identified as follows:%
\begin{equation}
F\simeq\left\{
\begin{array}
[c]{ccc}%
(R_{L}/R_{T})\Delta\rho gD^{2}/\eta_{2} &  & \eta_{2}\gg\eta_{1}\\
\eta_{1}VR_{T}R_{L}/D &  & \eta_{2}<\eta_{1}%
\end{array}
\right.  \label{e10}%
\end{equation}
with the corresponding velocity laws:%
\begin{equation}
V\simeq\left\{
\begin{array}
[c]{ccc}%
(R_{L}/R_{T})\Delta\rho gD^{2}/\eta_{2} &  & \eta_{2}\gg\eta_{1}\\
\Delta\rho gD^{2}/\eta_{1} &  & \eta_{2}<\eta_{1}%
\end{array}
\right.  \label{e11}%
\end{equation}
The crucial condition for these scaling laws to appear is that the formation
of thin layers of the surrounding liquid between the cell plates and both
surfaces of the disk-shaped droplet (see the right side view in Fig.
\ref{f3}). Otherwise, the drop would directly contact with the cell plates
making contact lines, and the dynamics would be affected by the contact angle
hysteresis (CAH). In general, physical understanding of CAH is very limited
and CAH is sensitive to even a slight contamination of the surface. Thus, if a
special care is properly taken for the surface of cell walls, the study of
drag friction under the existence of the contact line would be a challenging
future problem.

It is worth mentioning that in the three dimensional case, in the limit of
highly viscous fluid drop, $\eta_{2}\ll\eta_{1}$, which corresponds to the
viscous friction acting on a solid sphere, is given by Eq. (\ref{e6}), in
which $\eta_{1}$ is absent in contrast with Eq. (\ref{e10}). This is again
because of the existence of the thin layers of the surrounding fluid whose
thickness is $h$. However, when $\eta_{1}$ is extremely larger than $\eta_{2}%
$, depending on the thickness of the layers, dissipation associated with the
viscous stress $\eta_{2}V/h$ becomes smaller than that with $\eta_{1}V/D$.
This case is discussed in Sec. \ref{s5}.

In Fig. \ref{f6}, Eq. (\ref{e11})\ for $\eta_{2}<\eta_{1}$ is directly
confirmed, and thus Eq. (\ref{e10})\ for $\eta_{2}<\eta_{1}$ are indirectly
confirmed, for the data represented by the filled symbols. However, the data
represented by open symbols do not follow the scaling law, which is the case
mentioned in the previous paragraph and discussed in the next section. Note
that (as clear from theoretical arguments above) Eqs. (\ref{e10}) and
(\ref{e11})\ for $\eta_{2}>\eta_{1}$ are physically the same as Eqs.
(\ref{e7}) and (\ref{e8}), respectively, which is shown experimentally in
\cite{EriSoftMat2011}.

\section{Non-linear viscous drag friction}

\label{s5}

When the dissipation in the thin film is dominant, the velocity law is given
by%
\begin{equation}
\eta_{2}V\simeq\gamma_{12}(D/\kappa_{12}^{-1})^{3}\text{ \ for \ }%
D>\kappa_{12}^{-1} \label{e12}%
\end{equation}
and corresponding friction law is non-linear in $V$:%
\begin{equation}
F\simeq(\eta_{2}V/\gamma_{12})^{1/3}\gamma_{12}R_{T}R_{L}/\kappa_{12}%
^{-1}\text{ \ for \ }D>\kappa_{12}^{-1} \label{e13}%
\end{equation}
with the capillary length $\kappa_{12}^{-1}=\sqrt{\gamma_{12}/\Delta\rho g}$
defined for the drop-bath interface whose interfacial energy is $\gamma_{12}$.
For $D<\kappa_{12}^{-1}$, the velocity and force are replaced by the following
expressions, respectively: $\eta_{2}V\simeq\gamma_{12}(D/\kappa_{12}^{-1}%
)^{6}$ and $F\simeq(\eta_{2}V/\gamma_{12})^{1/3}\gamma_{12}R_{T}R_{L}/D$. We
note here that the viscous friction forces including this nonlinear friction
and that given in Eq. (\ref{e13}) are relevant to the dynamics of emulsion,
foam, antifoam and soft gels
\cite{Sylvie,AnnLaureSylvieSM2009,DominiqueMicrogravity2015}, in particular,
nonlinear rheology of such systems
\cite{DenkovSoftMat2009,DurianPRL10,CloitreNP2011}.

Equations (\ref{e12}) and (\ref{e13}) are derived because Eq. (\ref{e8a}) is
replaced with%
\begin{equation}
\eta_{2}(V/h)^{2}R_{T}R_{L}h\simeq\Delta\rho gR_{T}R_{L}DV, \label{e14}%
\end{equation}
with $h$ the thickness of the thin layers of the surrounding liquid, and
because, on the basis of the scaling arguments, \cite{CapilaryText} the
thickness of the film is expected to be given by
\begin{equation}
h=k\kappa_{12}^{-1}(\eta_{2}V/\gamma_{DB})^{2/3} \label{e15}%
\end{equation}
The value numerical coefficient $k$ is estimated as 0.94 in a more rigorous
calculation in Refs. \cite{LandauLevich,Derjaguin1943}. With Eq. (\ref{e12}),
Eq. (\ref{e15}) can be expressed as%
\begin{equation}
h\simeq\kappa_{12}^{-1}(D/\kappa_{12}^{-1})^{2}\text{ \ for \ }D>\kappa
_{12}^{-1} \label{e16}%
\end{equation}
For $D<\kappa_{12}^{-1}$, the thickness is given by the expression
\cite{ParkHomsyHeleShaw1984} $h\simeq D(\eta_{2}V/\gamma_{12})^{2/3}$ $\simeq
D(D/\kappa_{12}^{-1})^{4}$, which leads to the above-mentioned velocity and
drag force for $D<\kappa_{12}^{-1}$.

\begin{figure}[ptb]
\includegraphics[width=0.45\textwidth]{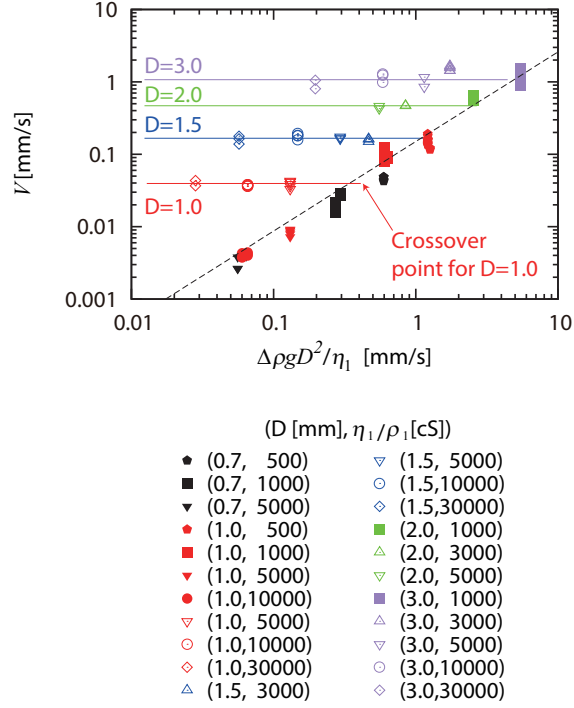}\caption{$V$ vs. $\Delta\rho
gD^{2}/\eta_{1}$. Data represented by the filled symbols are well on the
dashed line, confirming the scaling law given in Eq. (\ref{e11})\ for
$\eta_{2}<\eta_{1}$. The data including the data represented by the open
symbols are further discussed in Fig. \ref{f7}. Figures are adapted from Ref.
\cite{yahashi2016}.}%
\label{f6}%
\end{figure}

\begin{figure}[ptb]
\includegraphics[width=0.45\textwidth]{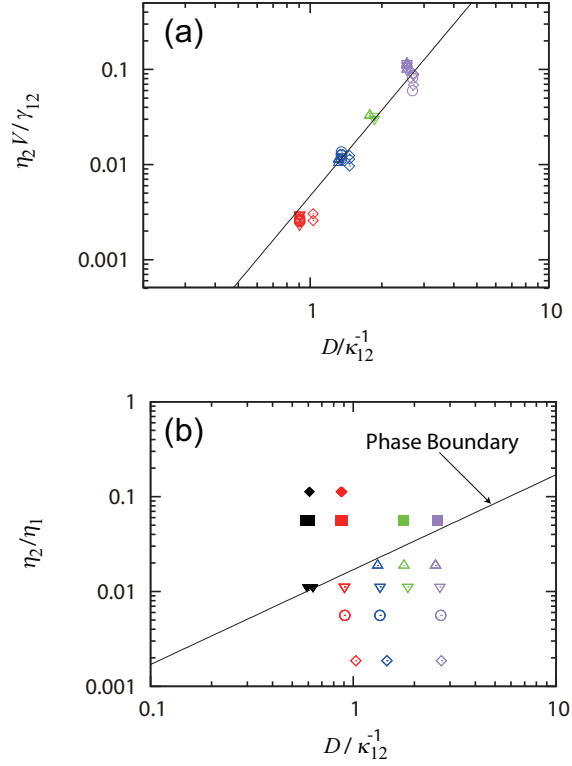}\caption{(a) $\eta_{1}%
V/\gamma_{12}$ vs. $D/\kappa_{12}^{-1}$ on a log-log scale. Data represented
by the open symbols are well on the line with slope 3, confirming the scaling
law given in Eq. (\ref{e12})\ for $D>\kappa_{12}^{-1}$. (b) $\eta_{2}/\eta
_{1}$ vs. $D/\kappa_{12}^{-1}$, confirming the crossover condition given in
Eq. (\ref{e17}); the line representing phase boundary separates the data
represented by the filled symbols from those by the open symbols. The symbols
are the same ones with those used in Fig. \ref{f6}. Figures are adapted from
Ref. \cite{yahashi2016}.}%
\label{f7}%
\end{figure}

The scaling law in Eq. (\ref{e12}) is directly confirmed, and thus Eq.
(\ref{e13}) is indirectly confirmed, in Fig. \ref{f7}(a), as explained in the
caption. The data used in Fig. \ref{f7}(a) are the ones plotted in Fig.
\ref{f6} in a different way by using the same open symbols. In Fig. \ref{f6},
these data are not on the dashed line but on horizontal lines; instead, the
data with the same value of $D$ are on the same horizontal line. This is
consistent with Eq. (\ref{e12}) because $\eta_{2}$, $\gamma_{12}$, and
$\kappa_{12}^{-1}$ are approximately fixed for the data, and in such a case
the value of $\eta_{2}V$ is classified by the value of $D$.

The condition for the crossover between linear and non-linear viscous
frictions, corresponding to that between Eq. (\ref{e11}) for $\eta_{2}%
<\eta_{1}$ and Eq. (\ref{e12}), can be determined by comparing the magnitude
of dissipations of the two regimes: $\eta_{1}(V/D)^{2}R_{T}R_{L}D\simeq
\eta_{2}(V/h)^{2}R_{T}R_{L}h$. This condition can be written as%
\begin{equation}
\eta_{2}/\eta_{1}\simeq D/\kappa_{12}^{-1} \label{e17}%
\end{equation}
This condition is confirmed in Fig. \ref{f7}(b), as explained in the caption,
by using the data in Fig. \ref{f6}. From this phase diagram, we expect, for
example, the crossover between the two regimes for the series of data obtained
for $D\simeq1$ mm (red symbols). This crossover can be directly confirmed in
Fig. \ref{f6}. The left three red (open) symbols are on the horizontal line,
whereas right two red (filled) symbols are on the dashed line. Note that red
filled circles and red filled inverse triangles are exceptional data: these
data are obtained when droplets go down nearly on the same trail more than
once; as a result, mixing of the two liquids comes into play, which increases
the viscosity of the thin layers sandwiched between the bubble and cell plates
and makes the velocity gradient in the thin films unfavorable (see Ref.
\cite{yahashi2016} for the details).

\begin{figure}[ptb]
\includegraphics[width=0.45\textwidth]{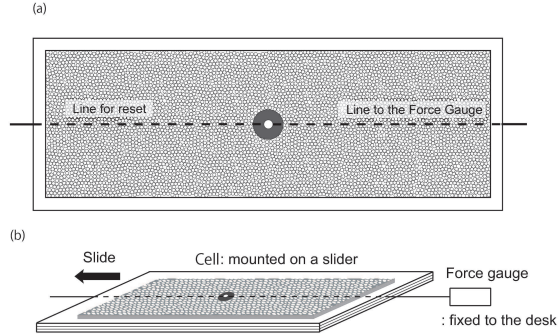}\caption{Setup for the
experiment for granular drag friction. (a) View of the cell from above. (b)
Slanted view of the setup. One layer of granular particles of average diameter
$d=2$ mm are packed in a two-dimensional cell made of transparent acrylic
plates. The diameter $2R$ of the obstacle is about ten times larger than $d$.
Reproduced from \cite{TakeharaPRL2014}.}%
\label{f8}%
\end{figure}

\begin{figure}[ptb]
\includegraphics[width=0.45\textwidth]{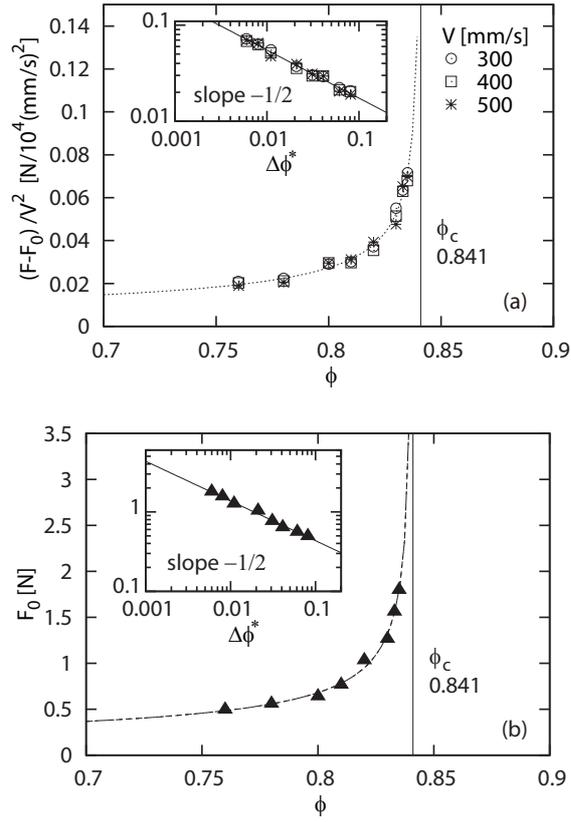}\caption{Divergence of the
dynamic (a) and statice (b) components of the granular drag force at the
jamming point $\phi=\phi_{c}$. Reproduced from \cite{TakeharaPRL2014}.}%
\label{f9}%
\end{figure}

Knowing that there are different scaling regimes for viscous drag friction in
the Hele-Shaw cell, a natural question to ask would be how the friction law
will change if the surrounding fluids are replaced with granular particles in
the cell. Such a study was performed by using the setup shown in Fig.
\ref{f8}. The cell filled with granular particles (diameter $d$ and density
$\rho$) can be moved at a constant velocity $V$ and the drag force $F$ acting
on the obstacle of radius $R$ can be monitored by the force gauge by virtue of
a non-extensible strong fishing line connecting the obstacle and the force
gauge. As a result, the drag force $F$ is found
\cite{TakeharaPRL2014,okumura2016PhilMag} to be given (1) by the dynamic part
scaling as $V^{2}$ governed by the momentum transfer associated with
collisions between the obstacle and cluster of grains ahead of it and (2) by
the static part $F_{0}$ governed by the friction force (friction coefficient
$\mu$) acting on the cluster from the bottom plate of the cell:%

\begin{align}
F-F_{0}  &  \simeq\rho(\phi_{c}-\phi)^{-1/2}R^{3}V^{2}/d\\
F_{0}  &  \simeq\mu\rho gR^{2}(\phi_{c}-\phi)^{-1/2}d
\end{align}
where $\phi$ is the packing fraction and $\phi_{c}$ is a critical density at
which the drag force diverges. This divergence, related to dynamic jamming of
grains ahead of the obstacle \cite{LiuNagel,bi2011jamming}, is clearly
confirmed in Fig. \ref{f9}. Interestingly, the value of $\phi_{c}\simeq0.841$
obtained from the experiment exactly matches the static jamming point in two
dimension \cite{NagelPRL02,NagelPRE03}. The background of this study is
briefly summarized as follows. The dynamic jamming has actively been explored
\cite{olsson2007critical,PouliquenPRL2011,MatthieuPNAS2012,JaegerEPL2013}.
Experiments on soft colloids \cite{DurianPRL10,CloitreNP2011} have
demonstrated good agreements with a phenomenological theory
\cite{vanHeckePRL2010}. However, as for granular systems, simulations and a
scaling phenomenology revealed critical behaviors different from those of soft
colloids \cite{HatanoJPSJ08,HayakawaPTP09,otsuki2012Progress}. Experimentally,
at high velocities ($\gtrsim100$ mm/s) and high densities, a force component
scaling with velocity squared has been reported through impact experiments
\cite{BehringerPRL2012,Goldman08,KatsuragiNM}, while different velocity
dependences have been reported in particular at much slower velocities
\cite{Wieghardt75,Schiffer1999,BehringerNature03,Chehata03,ReichhardtPRL03,DroccoPRL05,Schiffer2008,DauchotPRL09,EvelynePRE2013}%
. The divergence of the drag force at high velocities discussed above has been
examined in recent numerical studies \cite{takada2015drag,SeguinEPJE2016}.

\section{Drop Coalescence}

\begin{figure}[ptb]
\includegraphics[width=0.6\textwidth]{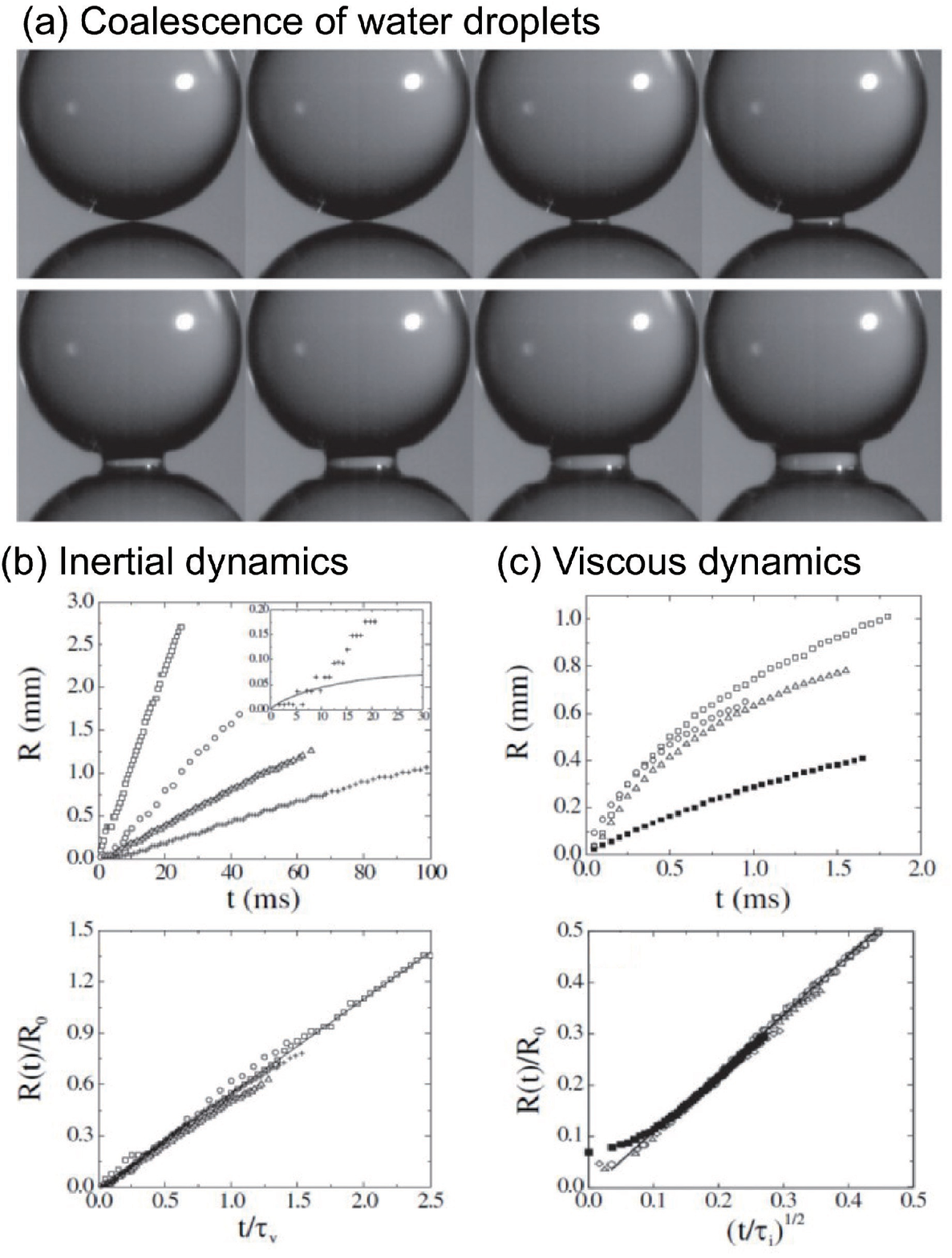}\caption{Three dimensional
coalescence. (a) Coalescence of water droplets captured by a high-speed camera
(11,200 frame per second). (b) Inertial dynamics. The raw data in the top
panel are collapse onto a single master curve in the bottom, by virtue of
rescaling of both axes according to Eq. (\ref{eq1}). (c) Viscous dynamics.
Another example of data collapse is shown, confirming Eq. (\ref{eq1}). Taken
from Ref. \cite{AartsLekkerkerkerGuoWegdamBonn2005}\copyright 2005 with
permission from APS.}%
\label{Fig1}%
\end{figure}

When a rain drop falling on the surface of a puddle, the rain drop coalesces
into the puddle. This is because the area of the liquid-air interface is made
smaller by the process. Figure \ref{Fig1}(a) shows a series of snapshots of
such a coalescence, but with minimizing the "falling velocity": the
coalescence is here initiated by touching the two droplets that are held at
the tips of pipets with a speed negligible for the coalescence dynamics
\cite{AartsLekkerkerkerGuoWegdamBonn2005}. From the figure, we see the
formation of a neck between the tips of the droplets that grows with time $t$.

Theoretical prediction was made analytically by Eggers and co-workers in 1999
\cite{Eggers1999}, and here we reproduce the result at the level of scaling
laws. The driving force is surface tension, which is opposed by inertial and
viscous forces. We estimate changes of the corresponding three energies per
unit time. The change of surface energy, $dE_{c}/dt\simeq d(\gamma_{1}%
r^{2})/dt\simeq\gamma_{12}r^{2}/t$, with the neck size $r$ and the surface
energy (per unit area) $\gamma_{1}$, which can be expressed as $\gamma_{12}$
(the interfacial energy between the drop and the surrounding fluid, i.e.,
air). This is because when a neck with radius $r$ is formed, the total area
the two droplets is decreased by an amount of order of $r^{2}$. This energy
will be lost by viscous dissipation and/or changed into kinetic energy.

In the small $r$ limit, viscous dissipation dominates because the volume of
moving region is small. In such a case, the two droplets contact with each
other nearly by a point, implying that there is only a single length scale for
the moving region, which is $r$. As a result, viscous dissipation per time is
given as $dE_{v}/dt\simeq\eta_{1}(V/r)^{2}r^{3}$, where $\eta_{1}$ is
viscosity of the liquid drop and $V\simeq r/t$ is the growth rate of the neck
size. This is because viscous dissipation per unit time and per unit volume
scales as $\eta_{1}(V/r)^{2}$ and the volume of dissipation scales as $r^{3}$.

When $r$ gets larger, kinetic energy comes into play. Since the length of neck
scales as $r^{2}/R$ from a geometrical consideration, kinetic-energy increase
per unit time scales as $dE_{i}/dt\simeq\rho_{1}r^{2}(r^{2}/R)V^{2}/t$.

From the above arguments, at short times surface energy is lost in viscous
dissipation: $dE_{c}/dt\simeq dE_{v}/dt$, that is, $r\simeq\gamma_{12}%
t/\eta_{1}$. In fact, Eggers and co-workers predicted an extra correction term
proportional to $\log t$, which has not been confirmed experimentally. In the
same way, at later times, surface energy is changed into kinetic energy:
$dE_{c}/dt\simeq dE_{i}/dt$, that is, $r\simeq(\gamma_{12}R/\rho_{1}%
)^{1/4}t^{1/2}$. The two scaling relations, one for the initial viscous regime
and another for the later inertial regime can be expressed in the following
dimensionless forms:%
\begin{equation}
r/R\simeq\left\{
\begin{array}
[c]{cc}%
t/\tau_{v} & \text{ for viscous regime }(t<\tau)\\
(t/\tau_{i})^{1/2} & \text{ for inertial regime }(t>\tau)
\end{array}
\right.  \label{eq1}%
\end{equation}
with $\tau_{v}=\eta_{1}R/\gamma_{12}$, $\tau_{i}=(\rho_{1}R^{3}/\gamma
_{12})^{1/2}$, and $\tau=\tau_{v}^{2}/\tau_{i}$.

An example of experimental confirmation of Eq. (\ref{eq1}) is given in Fig.
\ref{Fig1}(b) and (c), in which $r$ and $R$ are replaced with $R$ and $R_{0}$,
respectively. The top plots in (b) and (c) show the raw data in the two
regimes. The bottom plots show that the data are well collapsed onto the
master curves by rescaling of the axes, which confirms Eq. (\ref{eq1}).

\begin{figure}[ptb]
\includegraphics[width=0.45\textwidth]{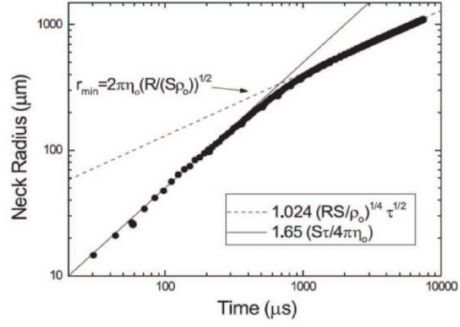}\caption{Purely
two-dimensional coalescence. (a) Coalescence of dodecane droplets floating on
water captured by a high-speed camera (the time interval between frames is 748
$\mu$s). The gravity points in the direction perpendicular to the paper. The
right illustration is the overview of two thin \textquotedblleft liquid
lens\textquotedblright\ floating on the horizontal surface of a bath of water,
which are represented by two circles contacting with each other at the center.
In the left, magnified snapshots of the central part are arranged in time
sequence (from left to right) showing the neck growth with time. (b) Scaling
crossover between the viscous and inertial regimes, confirming Eq.
(\ref{eq1}). Taken from Ref. \cite{BurtonTaborek2007}\copyright 2007 with
permission from APS.}%
\label{Fig2}%
\end{figure}

The crossover between the two regimes in Eq. (\ref{eq1}) is experimentally
confirmed in Fig. \ref{Fig2} \cite{BurtonTaborek2007}, in which $(r,\gamma
_{12},\eta_{1},\rho_{1},t)$ is replaced with $(r_{\min},S,\eta_{0},\rho
_{0},\tau)$. Here, this crossover is confirmed in a different system. When a
liquid drop of dodecane is placed on the surface of bath water, the drop takes
a thin lens shape. The coalescence of two of such floating droplets are
captured in Fig. \ref{Fig2}(a). In this geometry, the coalescence is found to
be purely two-dimensional, meaning that the velocity gradient in the direction
of the thickness of the lens is practically negligible, for which Eq.
(\ref{eq1}) is valid. As a result, a clear scaling crossover of the
coalescence dynamics is shown in Fig. \ref{Fig2}(b). Note that various
crossovers of scaling regimes for a coalescence of a fluid drop surrounded by
another fluid in three-dimension are recently discussed
\cite{NatCommunNagel2014Coalesce}.

\begin{figure}[ptb]
\includegraphics[width=0.7\textwidth]{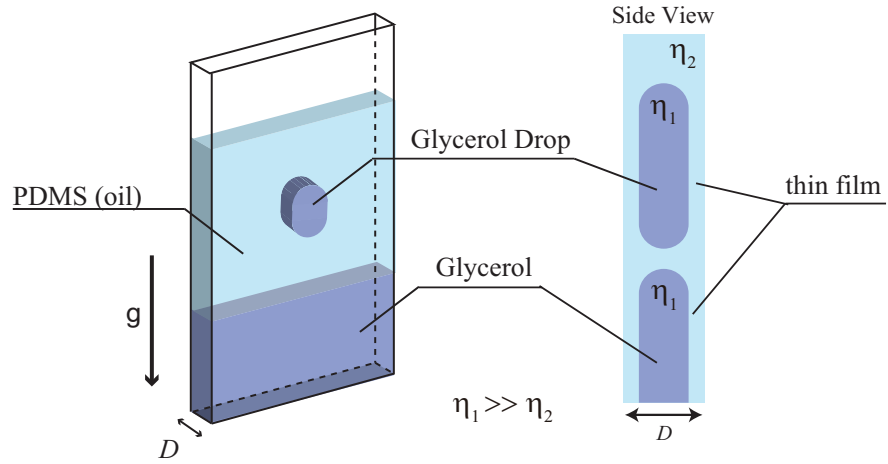}\caption{Experimental setup to
observe drop coalescence in a Hele-Shaw cell. The side view shows that the
existence of liquid thin films between the drop (or the lower bath of the same
liquid) and the cell walls. $\eta_{1}$ and $\eta_{2}$ respectively denote
viscosities of the liquid drop and the liquid surrounding the liquid drop. The
viscosity of the liquid in the lower bath phase is also $\eta_{1}$.}%
\label{Fig3}%
\end{figure}

Figure \ref{Fig3} shows an experimental setup to observe drop coalescence in a
Hele-Shaw cell \cite{YokotaPNAS2011}. The cell is held in the upright
position. First, the cell is half-filled by polydimethylsiloxane (PDMS).
Second, a glycerol aqueous solution is poured into the cell. Then, the heavier
glycerol solution replaces the lower PDMS phase. This replacement leaves thin
layers of PDMS in the lower phase, which are sandwiched by the bottom glycerol
phase and the cell plates (see the side view in Fig. \ref{Fig3}). This is
because the acrylic plates are completely wetted by the oil. After the
formation of two phases, a droplet of the glycerol solution is created in the
upper oil phase, which slowly goes down the oil phase to touch the
oil-glycerol interface, at which coalescence of the droplet and liquid bath of
the same liquid is initiated.

\begin{figure}[ptb]
\includegraphics[width=0.45\textwidth]{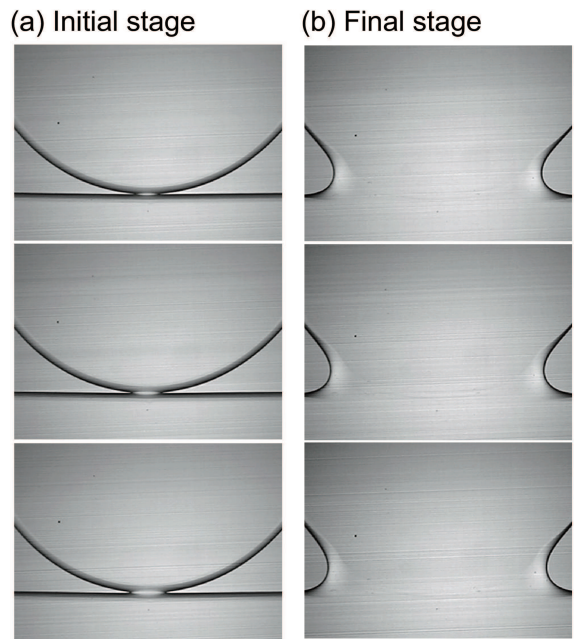}\caption{Coalescence in a
Hele-Shaw cell at short times (a) and at large times (a). The time intervals
between adjacent shots in (a) and (b) are 4 ms and 40 ms, respectively.
Adapted from \cite{YokotaPNAS2011}.}%
\label{Fig4}%
\end{figure}

Figure \ref{Fig4} shows snapshots at short times (initial stage) and those at
large times (final stage). These photographs are obtained by a high-speed
camera. The time interval between adjacent shots in Fig. \ref{Fig4}(b) are ten
times larger than that in Fig. \ref{Fig4}(a), indicating the dynamics slows
down significantly at large times.

\begin{figure}[ptb]
\includegraphics[width=0.45\textwidth]{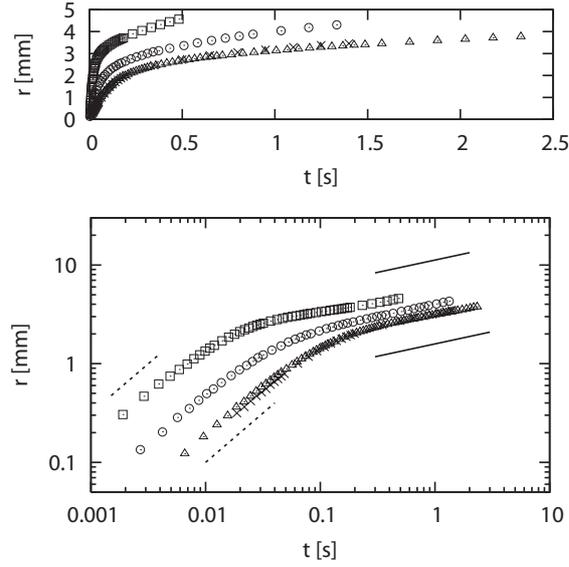}\caption{Neck size $r$ vs.
time $t$ for different parameters. The viscosity $\eta_{1}$ [mPa$\cdot$s] of
the drop, cell thickness $D$ [mm], and radius of the droplet $R$ [mm] for each
symbol are given as follows. $\square$: 62.9, 0.7, 5.62. ›: 289, 0.7,
5.56. $\triangle$: 888, 1.0, 4.13. $\times$: 964, 1.0, 4.32. (a) The raw data.
(b) The same data on a log-log scale, showing a clear scaling crossover. The
slopes that represent the initial and final dynamics are 1 and 1/4,
respectively. Adapted from \cite{YokotaPNAS2011}.}%
\label{Fig5}%
\end{figure}

Figure \ref{Fig5} shows two scaling regimes characterized by slope 1 (i.e.,
the neck size $r$ scales with $t$ at short times) and slope 1/4 (i.e., the
neck size $r$ scales with $t^{1/4}$ at large times), which helps us to
construct a theory. Before this, let us remark on experimental tricks
introduced in the present experiment. In this experiment, as explained in the
second paragraph above in which Fig. \ref{Fig3} is first mentioned, the oil
phase is deliberately introduced to remove the effect of contact line, which
is difficult to control: since the liquid disk is sandwiched by thin layers of
the oil, the disk has no direct contact with the cell walls. In addition, the
viscosity of oil $\eta_{2}$ is set to much smaller than that of the glycerol
solution $\eta_{1}$: the oil phase could play the role of air phase in the
three-dimensional counterpart.

Considering these points, we see that the dynamics of the initial regime could
be the same as the three-dimensional initial dynamics, which is characterized
by the slope 1. This is because at short times when $r$ is much smaller than
the cell thickness $D$ the dynamics could become independent of the existence
of the cell walls: the effect of confinement by the cell walls is practically negligible.

On the contrary, the dynamics at the final stage characterized by slope 1/4
reflects the effect of confinement of wall as discussed below. In the present
confined case, we have good reasons to make the following replacements from
the three-dimensional viscous counterpart: (1) The gain in surface energy per
unit time should be changed from $dE_{c}/dt\simeq d(\gamma_{12}r^{2})/dt$ to
$dE_{c}/dt\simeq d(\gamma_{12}rD)/dt$. (2) The viscous dissipation per unit
time should be changed from $dE_{v}/dt\simeq\eta_{1}(V/r)^{2}r^{3}$ to
$dE_{v}/dt\simeq\eta_{1}(V/D)^{2}rDd$ where $d$ is the height of the neck,
with which the volume of the neck is expressed as $rDd$. The first replacement
is rather obvious because the three-dimensional circular shape of the area
$\simeq r^{2}$ that disappears as a result of coalescence is here confined by
the wall to take a quasi-rectangular shape of area $\simeq rD$. The second
replacement is expected as long as $d\gtrsim D$, in which case the dominant
viscous dissipation per unit time and unit volume scales as $\eta_{1}%
(V/D)^{2}$ is localized in the neck volume $\simeq rDd$. In addition, from a
geometrical relation, $R^{2}=(R-d)^{2}+r^{2}$, the neck height $d$ is
expressed as $d\simeq r^{2}/R$ as long as $R\gtrsim d$. By balancing the
capillary energy gain with the viscous dissipation thus justified, we obtain
$r\simeq(\gamma_{12}RD^{2}t/\eta_{1})^{1/4}$, which leads to the desired slope 1/4.

\begin{figure}[ptb]
\includegraphics[width=0.45\textwidth]{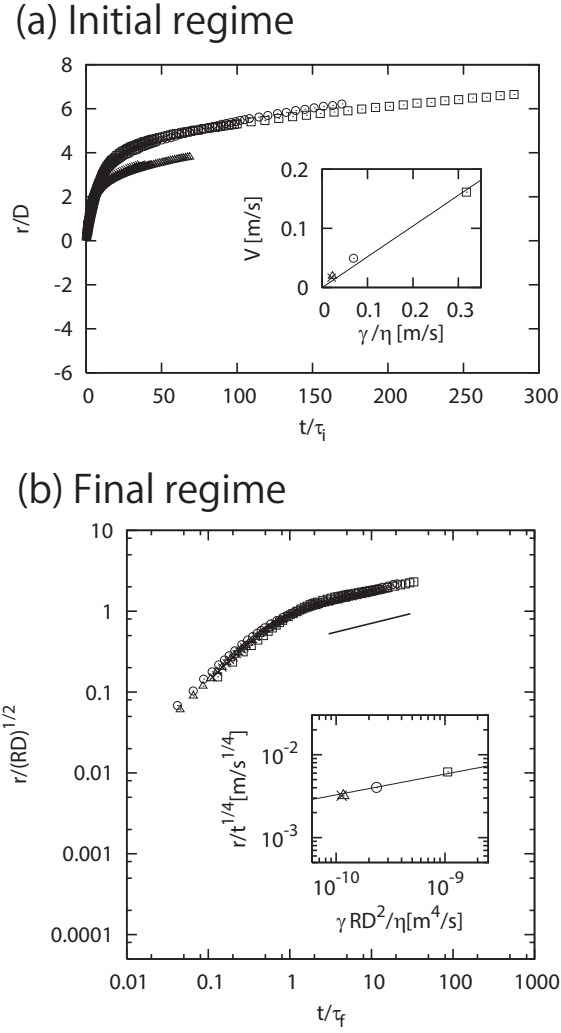}\caption{Confirmation of the
dimensional crossover expressed in Eqs. (\ref{eq2}) and (\ref{eq3}). (a)
Initial regime. (b) Final regime. In the main plots, data collapses are shown
for each regime. The insets show qualitative comparison of theory with
experiment. In the plots, $\tau_{I}$ and $\tau_{F}$ in Eqs. (\ref{eq2}) and
(\ref{eq3}) are expressed as $\tau_{i}$ and $\tau_{f}$, respectively. In
addition, $\gamma_{12}$ and $\eta_{1}$ are expressed by $\gamma$ and $\eta$,
respectively. Adapted from \cite{YokotaPNAS2011}.}%
\label{Fig6}%
\end{figure}

In summary, the dynamics exhibits a dimensional crossover from the short-time
three-dimensional one to the long-time quasi-two-dimensional one, expressed by
the following set of equations, which are confirmed in Fig. \ref{Fig6}.%

\begin{align}
r/D  &  \simeq t/\tau_{I}\text{ \ \ for }r<D\label{eq2}\\
r/\sqrt{RD}  &  \simeq(t/\tau_{F})^{1/4}\text{ \ \ for }\sqrt{RD}<r<R
\label{eq3}%
\end{align}
with $\tau_{I}=\eta_{1}D/\gamma_{12}$, and $\tau_{F}=\eta_{1}R/\gamma_{12}$.
Some remarks are as follows. (1) Equation (\ref{eq2}) is another dimensionless
form of the viscous regime given in Eq. (\ref{eq1}). (2) The condition,
$\sqrt{RD}<r<R$, for the final regime to be valid implies that this dynamics
is well observed when $D\ll R$ is satisfied, which is true for the
experimental data. (3) The final regime is predicted to start from $r$ and $t$
are larger than $\sqrt{RD}$ and $\tau_{F}$, respectively, which are well
confirmed in Fig. \ref{Fig6}(b).

There are a number of future problems for the quasi-two-dimensional
coalescence. One is to find an inertial regime since only viscous regimes have
been found. In the three-dimensional liquid drop coalescence in air, the
Ohnesorg number (Oh $\simeq\eta_{1}/\sqrt{\rho_{1}\gamma_{1}R}$) appears when
we discuss the crossover between the viscous and inertial regimes given in Eq.
(\ref{eq1}): we can express the crossover by the simple expression $r/R=$ Oh.
In the quasi-two dimensional case, the corresponding number will be useful to
look for different regimes including inertial regimes. Another interesting
problem would be to find partial coalescence in the quasi-two-dimensional
case, which has been discussed for a number of systems in three dimension
\cite{Mason1960PartialCoal,NatPhys2006PartialCoal,Bushi2015partial}.

\section{Film bursting in a Hele-Shaw cell}

\begin{figure}[ptb]
\includegraphics[width=0.45\textwidth]{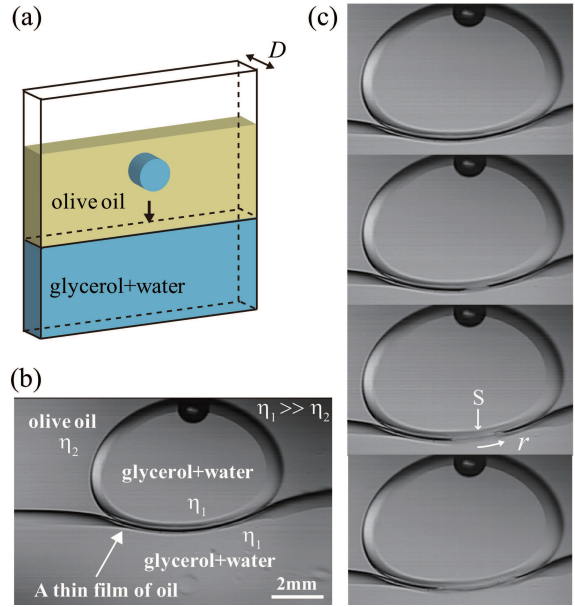}\caption{(a) Illustration of
the experiment. (b) A glycerol droplet sitting at the oil-glycerol interface
where a thin oil film underneath the droplet disallows coalescence typically
for a few minutes. The white bar stands for the cell thickness $D$ $(=2$ mm).
(c) The dynamics of coalescence: bursting of a thin film of oil between the
droplet and the bath. The snapshots are separated by 15/8000 seconds. The
length $r$ stands for the distance between the bursting tip and the point
(indicated as S) where the bursting started. }%
\label{Fig7}%
\end{figure}

When the coalescence experiment is performed with a different combination of
liquids, glycerol solution and olive oil, the mode of coalescence
significantly changes as shown in Fig. \ref{Fig7}, because thin film
(viscosity $\eta_{2}$) formed between the drop and bath of the same liquid
(viscosity $\eta_{1}$) is more stable in the present case (molecules contained
in olive oil may play a role similar to surfactants). This problem can be seen
as a problem of film bursting.

Bursting of liquid thin film is familiar to everybody as the rupture of soap
films, and is important in many fields
\cite{ThinLiquidFilmRMP97,Reiter2005,EliePRL07}. Although the fundamental
physics of the dynamics of bursting film is rather well and simply understood
\cite{CapilaryText}, in all previous studies the bursting hole has circular
symmetry, which is lost in the present case. Here, the problem is bursting of
a thin film in a confined geometry, which is relevant in many practical
situations where small amount of liquids has to be manipulated (e.g.,
microfluidics and biological applications).

When the circular symmetry is preserved, the bursting or dewetting frequently
proceeds with a constant speed and in such cases a rim is formed as in the
bursting of soap film in air
\cite{culick1960comments,Mysels1969bursting,frankel1969bursting} or in the
surrounding oil phase \cite{EtienneBurstingInOil} and in the bursting of
liquid film on a solid substrate (i.e. dewetting) \cite{DewetFrancoisePRL91};
when the rim is not formed the speed grows exponentially as observed in the
bursting of highly viscous films suspended in air
\cite{debregeas1995viscous,DebregeasGennesBrochard-Wyart1998}.

On the contrary, when the circular symmetry is lost, the bursting proceeds
with a constant speed but no rim is formed and the velocity $V$ is found
\cite{EriOkumura2010} to be a constant given by%
\begin{equation}
V\simeq\gamma_{12}/\eta_{1}%
\end{equation}
when $\eta_{1}\gtrsim\eta_{2}$. This scaling law can be derived by the balance
of energy per unit time:%
\begin{equation}
\gamma_{12}DV\simeq\eta_{1}(V/h)^{2}h^{2}D
\end{equation}
This scaling regime is significantly different from the three-dimensional case
because the dynamics proceeds with a constant speed although the rim is not formed.

\section{Conclusion}

In this review, we discussed viscous dynamics of bubbles in Hele-Shaw cells,
focusing on drainage and drag friction. We further explored the problem of
drag friction in a Hele-Shaw cell when the bubble is replaced by a viscous
drop or the surrounding liquid is replaced by granular medium to demonstrate
examples of nonlinear friction dynamics. In addition, viscous coalescence of a
liquid drop with a bath of the same liquid is also discussed, together with
film bursting problems. The collection of examples provided in this review is
not exhaustive but enough to show that the effects of confinement provided by
Hele-Shaw cells are worth studying further in the near future. The scaling
laws that emerge from these examples are relevant for industrial applications
dealing with drops and bubbles in confined spaces such as petroleum industry
\cite{HeleShawPetroleum2010} and microfluidic applications
\cite{SquiresQuake2005}, as well as for industrial and fundamental problems
associated with foams and microemulsions \cite{PhysicsFoams,Sylvie}. The
scaling laws in the Hele-Shaw geometry discussed here have been established
only by agreement between experiments and scaling phenomenologies, providing
future challenges for analytical and numerical studies.

\begin{acknowledgments}
This work was partly supported by Grant-in-Aid for Scientific Research (A)
(No. 24244066) of JSPS, Japan, and by ImPACT Program of Council for Science,
Technology and Innovation (Cabinet Office, Government of Japan).
\end{acknowledgments}


\begin{thebibliography}{99}                                                                                               %
\expandafter\ifx\csname url\endcsname\relax


\fi
\expandafter\ifx\csname urlprefix\endcsname\relax

\fi
\providecommand{\bibinfo}[2]{#2} \providecommand{\eprint}[2][]{\url{#2}}

\bibitem {DebregeasGennesBrochard-Wyart1998}%
\bibinfo{author}{Debr\'{e}geas, G.}, \bibinfo{author}{de~Gennes, P.-G.} \&
\bibinfo{author}{Brochard-Wyart, F.}
\newblock \bibinfo{title}{The life and death of "bare" viscous bubbles}.
\newblock \emph{\bibinfo{journal}{Science}} \textbf{\bibinfo{volume}{279}},
\bibinfo{pages}{1704--1707} (\bibinfo{year}{1998}).

\bibitem {EriOkumura2007}\bibinfo{author}{Eri, A.} \&
\bibinfo{author}{Okumura, K.}
\newblock \bibinfo{title}{Lifetime of a two-dimensional air bubble}.
\newblock \emph{\bibinfo{journal}{Phys. Rev. E}} \textbf{\bibinfo{volume}{76}}%
,  \bibinfo{pages}{060601(R)} (\bibinfo{year}{2007}).

\bibitem {EriSoftMat2011}\bibinfo{author}{Eri, A.} \&
\bibinfo{author}{Okumura, K.}
\newblock \bibinfo{title}{Viscous drag friction acting on a fluid drop confined
in between two plates confined in between two plates}.
\newblock \emph{\bibinfo{journal}{Soft Matter}} \textbf{\bibinfo{volume}{7}},
\bibinfo{pages}{5648} (\bibinfo{year}{2011}).

\bibitem {TAYLORSAFFMAN1959}\bibinfo{author}{Taylor, G.} \&
\bibinfo{author}{Saffman, P.~G.}
\newblock \bibinfo{title}{A note on the motion of bubbles in a hele-shaw cell
and porous medium}.
\newblock \emph{\bibinfo{journal}{Quarterly J. Mech. Applied Math.}}
\textbf{\bibinfo{volume}{12}}, \bibinfo{pages}{265--279}  (\bibinfo{year}{1959}).

\bibitem {Tanveer1986}\bibinfo{author}{Tanveer, S.}
\newblock \bibinfo{title}{The effect of surface tension on the shape of a
hele--shaw cell bubble}. \newblock \emph{\bibinfo{journal}{Phys. Fluids}}
\textbf{\bibinfo{volume}{29}},  \bibinfo{pages}{3537--3548} (\bibinfo{year}{1986}).

\bibitem {Bretherton}\bibinfo{author}{Bretherton, F.~P.}
\newblock \bibinfo{title}{The motion of long bubbles in tubes.}
\newblock \emph{\bibinfo{journal}{J. Fluid. Mech.}}
\textbf{\bibinfo{volume}{10}}, \bibinfo{pages}{166} (\bibinfo{year}{1961}).

\bibitem {Clanet2004}\bibinfo{author}{Clanet, C.},
\bibinfo{author}{H\'{e}raud, P.} \&  \bibinfo{author}{Searby, G.}
\newblock \bibinfo{title}{On the motion of bubbles in vertical tubes of
arbitrary cross-sections: Some complements to the dumitrescu-taylor problem}.
\newblock \emph{\bibinfo{journal}{J. Fluid Mech.}}
\textbf{\bibinfo{volume}{519}}, \bibinfo{pages}{359--376}  (\bibinfo{year}{2004}).

\bibitem {Eggers1997}\bibinfo{author}{Eggers, J.}
\newblock \bibinfo{title}{Nonlinear dynamics and breakup of free-surface
flows}. \newblock \emph{\bibinfo{journal}{Rev. Mod. Phys.}}
\textbf{\bibinfo{volume}{69}}, \bibinfo{pages}{865--930}  (\bibinfo{year}{1997}).

\bibitem {ConstantinDupontGoldsteinKadanoffShelleyZhou1993}%
\bibinfo{author}{Constantin, P.} \emph{et~al.}
\newblock \bibinfo{title}{Droplet breakup in a model of the hele-shaw cell}.
\newblock \emph{\bibinfo{journal}{Phys. Rev. E}} \textbf{\bibinfo{volume}{47}}%
,  \bibinfo{pages}{4169--4181} (\bibinfo{year}{1993}).

\bibitem {GoldsteinPesciShelley1995}\bibinfo{author}{Goldstein, R.~E.},
\bibinfo{author}{Pesci, A.~I.} \&  \bibinfo{author}{Shelley, M.~J.}
\newblock \bibinfo{title}{Attracting manifold for a viscous topology
transition}. \newblock \emph{\bibinfo{journal}{Phys. Rev. Lett.}}
\textbf{\bibinfo{volume}{75}}, \bibinfo{pages}{3665--3668}  (\bibinfo{year}{1995}).

\bibitem {Howell1999}\bibinfo{author}{Howell, P.~D.}
\newblock \bibinfo{title}{The draining of a two-dimensional bubble}.
\newblock \emph{\bibinfo{journal}{J. Eng. Math.}}
\textbf{\bibinfo{volume}{35}}, \bibinfo{pages}{251--272}  (\bibinfo{year}{1999}).

\bibitem {Maxworthy1986}\bibinfo{author}{Maxworthy, T.}
\newblock \bibinfo{title}{Bubble formation, motion and interaction in a
hele-shaw cell.} \newblock \emph{\bibinfo{journal}{J. Fluid Mech.}}
\textbf{\bibinfo{volume}{173}}, \bibinfo{pages}{95--114}  (\bibinfo{year}{1986}).

\bibitem {Kopf-SillHomsy1988}\bibinfo{author}{Kopf-Sill, A.~R.} \&
\bibinfo{author}{Homsy, G.~M.}
\newblock \bibinfo{title}{Bubble motion in a hele--shaw cell}.
\newblock \emph{\bibinfo{journal}{Phys. Fluids}} \textbf{\bibinfo{volume}{31}}%
,  \bibinfo{pages}{18--26} (\bibinfo{year}{1988}).

\bibitem {MaruvadaPark1996}\bibinfo{author}{Maruvada, S. R.~K.} \&
\bibinfo{author}{Park, C.-W.}
\newblock \bibinfo{title}{Retarded motion of bubbles in hele--shaw cells}.
\newblock \emph{\bibinfo{journal}{Phys. Fluids}} \textbf{\bibinfo{volume}{8}}%
,  \bibinfo{pages}{3229--3233} (\bibinfo{year}{1996}).

\bibitem {HeleShawPetroleum2010}\bibinfo{author}{Shad, S.},
\bibinfo{author}{Salarieh, M.},  \bibinfo{author}{Maini, B.} \&
\bibinfo{author}{Gates, I.~D.}
\newblock \bibinfo{title}{The velocity and shape of convected elongated liquid
drops in narrow gaps}.
\newblock \emph{\bibinfo{journal}{J. Petroleum Sci. Eng.}}
\textbf{\bibinfo{volume}{72}}, \bibinfo{pages}{67--77}  (\bibinfo{year}{2010}).

\bibitem {PascalEPL2002}\bibinfo{author}{Aussillous, P.} \&
\bibinfo{author}{Qu{\'e}r{\'e}, D.}
\newblock \bibinfo{title}{Bubbles creeping in a viscous liquid along a slightly
inclined plane}.
\newblock \emph{\bibinfo{journal}{EPL (Europhysics Letters)}}
\textbf{\bibinfo{volume}{59}}, \bibinfo{pages}{370} (\bibinfo{year}{2002}).

\bibitem {reyssat2014drops}\bibinfo{author}{Reyssat, E.}
\newblock \bibinfo{title}{Drops and bubbles in wedges}.
\newblock \emph{\bibinfo{journal}{J. Fluid. Mech.}}
\textbf{\bibinfo{volume}{748}}, \bibinfo{pages}{641--662}  (\bibinfo{year}{2014}).

\bibitem {CapsEPL2016}\bibinfo{author}{Dubois, C.},
\bibinfo{author}{Duchesne, A.} \&  \bibinfo{author}{Caps, H.}
\newblock \bibinfo{title}{Between inertia and viscous effects: Sliding bubbles
beneath an inclined plane}.
\newblock \emph{\bibinfo{journal}{EPL (Europhys. Lett.)}}
\textbf{\bibinfo{volume}{115}}, \bibinfo{pages}{44001}  (\bibinfo{year}{2016}).

\bibitem {StoneNM2007}\bibinfo{author}{Courbin, L.} \emph{et~al.}
\newblock \bibinfo{title}{Imbibition by polygonal spreading on microdecorated
surfaces}. \newblock \emph{\bibinfo{journal}{Nat. Mater.}}
\textbf{\bibinfo{volume}{6}},  \bibinfo{pages}{661--664} (\bibinfo{year}{2007}).

\bibitem {IshinoReyssatEPL2007}\bibinfo{author}{Ishino, C.},
\bibinfo{author}{Reyssat, M.},  \bibinfo{author}{Reyssat, E.},
\bibinfo{author}{Okumura, K.} \&  \bibinfo{author}{Qu\'{e}r\'{e}, D.}
\newblock \bibinfo{title}{Wicking within forests of micropillars}.
\newblock \emph{\bibinfo{journal}{Europhys. Lett.}}
\textbf{\bibinfo{volume}{79}}, \bibinfo{pages}{56005--(1--5)}  (\bibinfo{year}{2007}).

\bibitem {ObaraPRER2012}\bibinfo{author}{Obara, N.} \&
\bibinfo{author}{Okumura, K.}
\newblock \bibinfo{title}{Imbibition of a textured surface decorated by short
pillars with rounded edges}. \newblock \emph{\bibinfo{journal}{Phys. Rev. E}}
\textbf{\bibinfo{volume}{86}},  \bibinfo{pages}{020601(R)} (\bibinfo{year}{2012}).

\bibitem {TaniPlosOne2014}\bibinfo{author}{Tani, M.} \emph{et~al.}
\newblock \bibinfo{title}{Capillary rise on legs of a small animal and on
artificially textured surfaces mimicking them}.
\newblock \emph{\bibinfo{journal}{Plos One}} \textbf{\bibinfo{volume}{9}},
\bibinfo{pages}{e96813} (\bibinfo{year}{2014}).

\bibitem {Tani2015}\bibinfo{author}{Tani, M.}, \bibinfo{author}{Kawano, R.},
\bibinfo{author}{Kamiya, K.} \& \bibinfo{author}{Okumura, K.}
\newblock \bibinfo{title}{Towards combinatorial mixing devices without any
pumps by open-capillary channels: fundamentals and applications}.
\newblock \emph{\bibinfo{journal}{Sci. Rep.}} \textbf{\bibinfo{volume}{5}},
\bibinfo{pages}{10263--} (\bibinfo{year}{2015}).

\bibitem {DominicVellaImbibition2016}\bibinfo{author}{Gorce, J.-B.},
\bibinfo{author}{Hewitt, I.~J.} \&  \bibinfo{author}{Vella, D.}
\newblock \bibinfo{title}{Capillary imbibition into converging tubes: Beating
washburnfs law and the optimal imbibition of liquids}.
\newblock \emph{\bibinfo{journal}{Langmuir}} \textbf{\bibinfo{volume}{32}},
\bibinfo{pages}{1560--1567} (\bibinfo{year}{2016}).

\bibitem {Sylvie}\bibinfo{author}{Cantat, I.} \emph{et~al.}
\newblock \emph{\bibinfo{title}{Les mousses: structure et dynamique}}
(\bibinfo{publisher}{Belin, Paris}, \bibinfo{year}{2010}).

\bibitem {AnnLaureSylvieSM2009}\bibinfo{author}{Biance, A.-L.},
\bibinfo{author}{Cohen-Addad, S.} \&  \bibinfo{author}{H{\"o}hler, R.}
\newblock \bibinfo{title}{Topological transition dynamics in a strained bubble
cluster}. \newblock \emph{\bibinfo{journal}{Soft Matter}}
\textbf{\bibinfo{volume}{5}},  \bibinfo{pages}{4672--4679} (\bibinfo{year}{2009}).

\bibitem {DominiqueMicrogravity2015}\bibinfo{author}{Yazhgur, P.}
\emph{et~al.}
\newblock \bibinfo{title}{How antifoams act: a microgravity study}.
\newblock \emph{\bibinfo{journal}{npj Microgravity}}
\textbf{\bibinfo{volume}{1}} (\bibinfo{year}{2015}).

\bibitem {DenkovSoftMat2009}\bibinfo{author}{Denkov, N.~D.},
\bibinfo{author}{Tcholakova, S.},  \bibinfo{author}{Golemanov, K.},
\bibinfo{author}{Ananthpadmanabhan, K.} \&  \bibinfo{author}{Lips, A.}
\newblock \bibinfo{title}{The role of surfactant type and bubble surface
mobility in foam rheology}. \newblock \emph{\bibinfo{journal}{Soft Matter}}
\textbf{\bibinfo{volume}{5}},  \bibinfo{pages}{3389--3408} (\bibinfo{year}{2009}).

\bibitem {DurianPRL10}\bibinfo{author}{Nordstrom, K.} \emph{et~al.}
\newblock \bibinfo{title}{Microfluidic rheology of soft colloids above and
below jamming}. \newblock \emph{\bibinfo{journal}{Phys. Rev. Lett.}}
\textbf{\bibinfo{volume}{105}}, \bibinfo{pages}{175701}  (\bibinfo{year}{2010}).

\bibitem {CloitreNP2011}\bibinfo{author}{Seth, J.},
\bibinfo{author}{Mohan, L.},  \bibinfo{author}{Locatelli-Champagne, C.},
\bibinfo{author}{Cloitre, M.} \&  \bibinfo{author}{Bonnecaze, R.}
\newblock \bibinfo{title}{A micromechanical model to predict the flow of soft
particle glasses}. \newblock \emph{\bibinfo{journal}{Nature Mater.}}
\textbf{\bibinfo{volume}{10}}, \bibinfo{pages}{838--843}  (\bibinfo{year}{2011}).

\bibitem {CapilaryText}\bibinfo{author}{de~Gennes, P.-G.},
\bibinfo{author}{Brochard-Wyart, F.} \&  \bibinfo{author}{Qu\'{e}r\'{e}, D.}
\newblock \emph{\bibinfo{title}{Gouttes, Bulles, Perles et Ondes, 2nd. eds.}}
(\bibinfo{publisher}{Belin, Paris}, \bibinfo{year}{2005}).

\bibitem {LandauLevich}\bibinfo{author}{Landau, L.} \&
\bibinfo{author}{Levich, B.} \newblock \bibinfo{title}{Physicochim}.
\newblock \emph{\bibinfo{journal}{Acta. Physicochim (URSS)}}
\textbf{\bibinfo{volume}{17}}, \bibinfo{pages}{42} (\bibinfo{year}{1942}).

\bibitem {Derjaguin1943}\bibinfo{author}{Derhaguin, B.}
\newblock \bibinfo{title}{Physicochim}.
\newblock \emph{\bibinfo{journal}{Acta. Physicochim (URSS)}}
\textbf{\bibinfo{volume}{20}}, \bibinfo{pages}{349} (\bibinfo{year}{1943}).

\bibitem {ParkHomsyHeleShaw1984}\bibinfo{author}{Park, C.-W.} \&
\bibinfo{author}{Homsy, G.}
\newblock \bibinfo{title}{Two-phase displacement in hele shaw cells: theory}.
\newblock \emph{\bibinfo{journal}{J. Fluid Mech.}}
\textbf{\bibinfo{volume}{139}}, \bibinfo{pages}{291--308}  (\bibinfo{year}{1984}).

\bibitem {yahashi2016}\bibinfo{author}{Yahashi, M.},
\bibinfo{author}{Kimoto, N.} \&  \bibinfo{author}{Okumura, K.}
\newblock \bibinfo{title}{Scaling crossover in thin-film drag dynamics of fluid
drops in the hele-shaw cell}. \newblock \emph{\bibinfo{journal}{Sci. Rep.}}
\textbf{\bibinfo{volume}{6}},  \bibinfo{pages}{31395} (\bibinfo{year}{2016}).

\bibitem {TakeharaPRL2014}\bibinfo{author}{Takehara, Y.} \&
\bibinfo{author}{Okumura, K.}
\newblock \bibinfo{title}{High-velocity drag friction in granular media near
the jamming point}. \newblock \emph{\bibinfo{journal}{Phys. Rev. Lett.}}
\textbf{\bibinfo{volume}{112}}, \bibinfo{pages}{148001}  (\bibinfo{year}{2014}).

\bibitem {okumura2016PhilMag}\bibinfo{author}{Okumura, K.}
\newblock \bibinfo{title}{Simple views on different problems in physics: from
drag friction to tough biological materials}.
\newblock \emph{\bibinfo{journal}{Phil. Mag.}} \textbf{\bibinfo{volume}{96}},
\bibinfo{pages}{828--841} (\bibinfo{year}{2016}).

\bibitem {LiuNagel}\bibinfo{author}{Liu, A.~J.} \&
\bibinfo{author}{Nagel, S.~R.}
\newblock \bibinfo{title}{Nonlinear dynamics: Jamming is not just cool any
more}. \newblock \emph{\bibinfo{journal}{Nature}}
\textbf{\bibinfo{volume}{396}},  \bibinfo{pages}{21--22} (\bibinfo{year}{1998}).

\bibitem {bi2011jamming}\bibinfo{author}{Bi, D.},
\bibinfo{author}{Zhang, J.},  \bibinfo{author}{Chakraborty, B.} \&
\bibinfo{author}{Behringer, R.} \newblock \bibinfo{title}{Jamming by shear}.
\newblock \emph{\bibinfo{journal}{Nature}} \textbf{\bibinfo{volume}{480}},
\bibinfo{pages}{355--358} (\bibinfo{year}{2011}).

\bibitem {NagelPRL02}\bibinfo{author}{O'Hern, C.},
\bibinfo{author}{Langer, S.},  \bibinfo{author}{Liu, A.} \&
\bibinfo{author}{Nagel, S.}
\newblock \bibinfo{title}{Random packings of frictionless particles}.
\newblock \emph{\bibinfo{journal}{Phys. Rev. Lett.}}
\textbf{\bibinfo{volume}{88}}, \bibinfo{pages}{75507} (\bibinfo{year}{2002}).

\bibitem {NagelPRE03}\bibinfo{author}{O'Hern, C.},
\bibinfo{author}{Silbert, L.},  \bibinfo{author}{Liu, A.} \&
\bibinfo{author}{Nagel, S.}
\newblock \bibinfo{title}{Jamming at zero temperature and zero applied stress:
The epitome of disorder}. \newblock \emph{\bibinfo{journal}{Phys. Rev. E}}
\textbf{\bibinfo{volume}{68}},  \bibinfo{pages}{011306} (\bibinfo{year}{2003}).

\bibitem {olsson2007critical}\bibinfo{author}{Olsson, P.} \&
\bibinfo{author}{Teitel, S.}
\newblock \bibinfo{title}{Critical scaling of shear viscosity at the jamming
transition}. \newblock \emph{\bibinfo{journal}{Phys. Rev. Lett.}}
\textbf{\bibinfo{volume}{99}}, \bibinfo{pages}{178001}  (\bibinfo{year}{2007}).

\bibitem {PouliquenPRL2011}\bibinfo{author}{Boyer, F.},
\bibinfo{author}{Guazzelli, E.} \&  \bibinfo{author}{Pouliquen, O.}
\newblock \bibinfo{title}{Unifying suspension and granular rheology}.
\newblock \emph{\bibinfo{journal}{Phys. Rev. Lett.}}
\textbf{\bibinfo{volume}{107}}, \bibinfo{pages}{188301}  (\bibinfo{year}{2011}).

\bibitem {MatthieuPNAS2012}\bibinfo{author}{Lerner, E.},
\bibinfo{author}{D{\"u}ring, G.} \&  \bibinfo{author}{Wyart, M.}
\newblock \bibinfo{title}{A unified framework for non-brownian suspension flows
and soft amorphous solids}.
\newblock \emph{\bibinfo{journal}{Proc. Nat. Acad. Sci. (U.S.A.)}}
\textbf{\bibinfo{volume}{109}}, \bibinfo{pages}{4798--4803}  (\bibinfo{year}{2012}).

\bibitem {JaegerEPL2013}\bibinfo{author}{Waitukaitis, S.},
\bibinfo{author}{Roth, L.},  \bibinfo{author}{Vitelli, V.} \&
\bibinfo{author}{Jaeger, H.}
\newblock \bibinfo{title}{Dynamic jamming fronts}.
\newblock \emph{\bibinfo{journal}{EPL (Europhys. Lett.)}}
\textbf{\bibinfo{volume}{102}}, \bibinfo{pages}{44001}  (\bibinfo{year}{2013}).

\bibitem {vanHeckePRL2010}\bibinfo{author}{Tighe, B.~P.},
\bibinfo{author}{Woldhuis, E.},  \bibinfo{author}{Remmers, J. J.~C.},
\bibinfo{author}{van Saarloos, W.} \&  \bibinfo{author}{van Hecke, M.}
\newblock \bibinfo{title}{Model for the scaling of stresses and fluctuations in
flows near jamming}. \newblock \emph{\bibinfo{journal}{Phys. Rev. Lett.}}
\textbf{\bibinfo{volume}{105}}, \bibinfo{pages}{088303}  (\bibinfo{year}{2010}).

\bibitem {HatanoJPSJ08}\bibinfo{author}{Hatano, T.}
\newblock \bibinfo{title}{Scaling properties of granular rheology near the
jamming transition}. \newblock \emph{\bibinfo{journal}{J. Phys. Soc. Jpn.}}
\textbf{\bibinfo{volume}{77}}, \bibinfo{pages}{123002}  (\bibinfo{year}{2008}).

\bibitem {HayakawaPTP09}\bibinfo{author}{Otsuki, M.} \&
\bibinfo{author}{Hayakawa, H.}
\newblock \bibinfo{title}{Universal scaling for the jamming transition}.
\newblock \emph{\bibinfo{journal}{Prog. Theor. Phys.}}
\textbf{\bibinfo{volume}{121}}, \bibinfo{pages}{647--655}  (\bibinfo{year}{2009}).

\bibitem {otsuki2012Progress}\bibinfo{author}{Otsuki, M.} \&
\bibinfo{author}{Hayakawa, H.}
\newblock \bibinfo{title}{Rheology of sheared granular particles near jamming
transition}. \newblock \emph{\bibinfo{journal}{Prog. Theor. Phys. Suppl.}}
\textbf{\bibinfo{volume}{195}}, \bibinfo{pages}{129--138}  (\bibinfo{year}{2012}).

\bibitem {BehringerPRL2012}\bibinfo{author}{Clark, A.~H.},
\bibinfo{author}{Kondic, L.} \&  \bibinfo{author}{Behringer, R.~P.}
\newblock \bibinfo{title}{Particle scale dynamics in granular impact}.
\newblock \emph{\bibinfo{journal}{Phys. Rev. Lett.}}
\textbf{\bibinfo{volume}{109}}, \bibinfo{pages}{238302}  (\bibinfo{year}{2012}).

\bibitem {Goldman08}\bibinfo{author}{Goldman, D.~I.} \&
\bibinfo{author}{Umbanhowar, P.}
\newblock \emph{\bibinfo{journal}{Phys. Rev. E}} \textbf{\bibinfo{volume}{77}}%
,  \bibinfo{pages}{021308--} (\bibinfo{year}{2008}).

\bibitem {KatsuragiNM}\bibinfo{author}{Katsuragi, H.} \&
\bibinfo{author}{Durian, D.~J.}
\newblock \emph{\bibinfo{journal}{Nature Phys.}} \textbf{\bibinfo{volume}{3}}%
,  \bibinfo{pages}{420--} (\bibinfo{year}{2007}).

\bibitem {Wieghardt75}\bibinfo{author}{Wieghardt, K.}
\newblock \emph{\bibinfo{journal}{Annu. Rev. Fluid. Mech.}}
\textbf{\bibinfo{volume}{7}}, \bibinfo{pages}{89--} (\bibinfo{year}{1975}).

\bibitem {Schiffer1999}\bibinfo{author}{Albert, R.},
\bibinfo{author}{Pfeifer, M.},  \bibinfo{author}{Barab{\'a}si, A.-L.} \&
\bibinfo{author}{Schiffer, P.}
\newblock \bibinfo{title}{Slow drag in a granular medium}.
\newblock \emph{\bibinfo{journal}{Phys. Rev. Lett.}}
\textbf{\bibinfo{volume}{82}}, \bibinfo{pages}{205} (\bibinfo{year}{1999}).

\bibitem {BehringerNature03}\bibinfo{author}{Hartley, R.~R.} \&
\bibinfo{author}{Behringer, R.~P.} \newblock \emph{\bibinfo{journal}{Nature}}
\textbf{\bibinfo{volume}{421}},  \bibinfo{pages}{928--} (\bibinfo{year}{2003}).

\bibitem {Chehata03}\bibinfo{author}{Chehata, D.}, \bibinfo{author}{Zenit, R.}
\&  \bibinfo{author}{Wassgren, C.~R.}
\newblock \emph{\bibinfo{journal}{Phys. Fluids}} \textbf{\bibinfo{volume}{15}}%
,  \bibinfo{pages}{1622--} (\bibinfo{year}{2003}).

\bibitem {ReichhardtPRL03}\bibinfo{author}{Hastings, M.~B.},
\bibinfo{author}{Olson~Reichhardt, C.~J.} \&  \bibinfo{author}{Reichhardt, C.}
\newblock \bibinfo{title}{Depinning by fracture in a glassy background}.
\newblock \emph{\bibinfo{journal}{Phys. Rev. Lett.}}
\textbf{\bibinfo{volume}{90}}, \bibinfo{pages}{098302}  (\bibinfo{year}{2003}).

\bibitem {DroccoPRL05}\bibinfo{author}{Drocco, J.~A.},
\bibinfo{author}{Hastings, M.~B.},  \bibinfo{author}{Reichhardt, C. J.~O.} \&
\bibinfo{author}{Reichhardt, C.}
\newblock \emph{\bibinfo{journal}{Phys. Rev. Lett.}}
\textbf{\bibinfo{volume}{95}}, \bibinfo{pages}{088001--}  (\bibinfo{year}{2005}).

\bibitem {Schiffer2008}\bibinfo{author}{Costantino, D.~J.} \emph{et~al.}
\newblock \bibinfo{title}{Starting to move through a granular medium}.
\newblock \emph{\bibinfo{journal}{Phys. Rev. Lett.}}
\textbf{\bibinfo{volume}{101}}, \bibinfo{pages}{108001}  (\bibinfo{year}{2008}).

\bibitem {DauchotPRL09}\bibinfo{author}{Candelier, R.} \&
\bibinfo{author}{Dauchot, O.}
\newblock \bibinfo{title}{Creep motion of an intruder within a granular glass
close to jamming}. \newblock \emph{\bibinfo{journal}{Phys. Rev. Lett.}}
\textbf{\bibinfo{volume}{103}}, \bibinfo{pages}{128001}  (\bibinfo{year}{2009}).

\bibitem {EvelynePRE2013}\bibinfo{author}{Kolb, E.},
\bibinfo{author}{Cixous, P.},  \bibinfo{author}{Gaudouen, N.} \&
\bibinfo{author}{Darnige, T.}
\newblock \bibinfo{title}{Rigid intruder inside a two-dimensional dense
granular flow: Drag force and cavity formation}.
\newblock \emph{\bibinfo{journal}{Phys. Rev. E}} \textbf{\bibinfo{volume}{87}}%
,  \bibinfo{pages}{032207} (\bibinfo{year}{2013}).

\bibitem {takada2015drag}\bibinfo{author}{Takada, S.} \&
\bibinfo{author}{Hayakawa, H.}
\newblock \bibinfo{title}{Drag law of two dimensional granular fluids}.
\newblock \emph{\bibinfo{journal}{J. Eng. Mech.}} \bibinfo{pages}{doi:
10.1061/(ASCE)EM.1943--7889.0001054} (\bibinfo{year}{2016}).

\bibitem {SeguinEPJE2016}\bibinfo{author}{Seguin, A.},
\bibinfo{author}{Lefebvre-Lepot, A.},  \bibinfo{author}{Faure, S.} \&
\bibinfo{author}{Gondret, P.}
\newblock \bibinfo{title}{Clustering and flow around a sphere moving into a
grain cloud}. \newblock \emph{\bibinfo{journal}{Eur. Phys. J. E}}
\textbf{\bibinfo{volume}{39}}, \bibinfo{pages}{63--} (\bibinfo{year}{2016}).

\bibitem {AartsLekkerkerkerGuoWegdamBonn2005}%
\bibinfo{author}{Aarts, D. G. A.~L.},
\bibinfo{author}{Lekkerkerker, H. N.~W.},  \bibinfo{author}{Guo, H.},
\bibinfo{author}{Wegdam, G.~H.} \&  \bibinfo{author}{Bonn, D.}
\newblock \bibinfo{title}{Hydrodynamics of droplet coalescence}.
\newblock \emph{\bibinfo{journal}{Phys. Rev. Lett.}}
\textbf{\bibinfo{volume}{95}}, \bibinfo{pages}{164503}  (\bibinfo{year}{2005}).

\bibitem {Eggers1999}\bibinfo{author}{Eggers, J.},
\bibinfo{author}{Lister, J.} \&  \bibinfo{author}{Stone, H.}
\newblock \bibinfo{title}{Coalescence of liquid drops}.
\newblock \emph{\bibinfo{journal}{J. Fluid Mech.}}
\textbf{\bibinfo{volume}{401}}, \bibinfo{pages}{293--310}  (\bibinfo{year}{1999}).

\bibitem {BurtonTaborek2007}\bibinfo{author}{Burton, J.~C.} \&
\bibinfo{author}{Taborek, P.}
\newblock \bibinfo{title}{Role of dimensionality and axisymmetry in fluid
pinch-off and coalescence}.
\newblock \emph{\bibinfo{journal}{Phys. Rev. Lett.}}
\textbf{\bibinfo{volume}{98}}, \bibinfo{pages}{224502}  (\bibinfo{year}{2007}).

\bibitem {NatCommunNagel2014Coalesce}\bibinfo{author}{Paulsen, J.~D.},
\bibinfo{author}{Carmigniani, R.},  \bibinfo{author}{Kannan, A.},
\bibinfo{author}{Burton, J.~C.} \&  \bibinfo{author}{Nagel, S.~R.}
\newblock \bibinfo{title}{Coalescence of bubbles and drops in an outer fluid}.
\newblock \emph{\bibinfo{journal}{Nature Commun.}}
\textbf{\bibinfo{volume}{5}}  (\bibinfo{year}{2014}).

\bibitem {YokotaPNAS2011}\bibinfo{author}{Yokota, M.} \&
\bibinfo{author}{Okumura, K.}
\newblock \bibinfo{title}{Dimensional crossover in the coalescence dynamics of
viscous drops confined in between two plates}.
\newblock \emph{\bibinfo{journal}{Proc. Nat. Acad. Sci. (U.S.A.)}}
\textbf{\bibinfo{volume}{108}}, \bibinfo{pages}{6395--6398; In this issue,
PNAS, 108 (2011) 6337.} (\bibinfo{year}{2011}).

\bibitem {Mason1960PartialCoal}\bibinfo{author}{Charles, G.~E.} \&
\bibinfo{author}{Mason, S.~G.}
\newblock \bibinfo{title}{The mechanism of partial coalescence of liquid drops
at liquid/liquid interfaces}.
\newblock \emph{\bibinfo{journal}{Journal of Colloid Science}}
\textbf{\bibinfo{volume}{15}}, \bibinfo{pages}{105--122}  (\bibinfo{year}{1960}).

\bibitem {NatPhys2006PartialCoal}\bibinfo{author}{Blanchette, F.} \&
\bibinfo{author}{Bigioni, T.~P.}
\newblock \bibinfo{title}{Partial coalescence of drops at liquid interfaces}.
\newblock \emph{\bibinfo{journal}{Nature Physics}}
\textbf{\bibinfo{volume}{2}}, \bibinfo{pages}{254--257}  (\bibinfo{year}{2006}).

\bibitem {Bushi2015partial}\bibinfo{author}{Pucci, G.},
\bibinfo{author}{Harris, D.~M.} \&  \bibinfo{author}{Bush, J.~W.}
\newblock \bibinfo{title}{Partial coalescence of soap bubbles}.
\newblock \emph{\bibinfo{journal}{Phys. Fluids}} \textbf{\bibinfo{volume}{27}}%
,  \bibinfo{pages}{061704} (\bibinfo{year}{2015}).

\bibitem {ThinLiquidFilmRMP97}\bibinfo{author}{Oron, A.},
\bibinfo{author}{Davis, S.~H.} \&  \bibinfo{author}{Bankoff, S.~G.}
\newblock \bibinfo{title}{Long-scale evolution of thin liquid films}.
\newblock \emph{\bibinfo{journal}{Rev. Mod. Phys.}}
\textbf{\bibinfo{volume}{69}}, \bibinfo{pages}{931--980}  (\bibinfo{year}{1997}).

\bibitem {Reiter2005}\bibinfo{author}{Reiter, G.} \emph{et~al.}
\newblock \bibinfo{title}{Residual stresses in thin polymer films cause rupture
and dominate early stages of dewetting}.
\newblock \emph{\bibinfo{journal}{Nature Mater.}} \textbf{\bibinfo{volume}{4}}%
,  \bibinfo{pages}{754--758} (\bibinfo{year}{2005}).

\bibitem {EliePRL07}\bibinfo{author}{Damman, P.} \emph{et~al.}
\newblock \bibinfo{title}{Relaxation of residual stress and reentanglement of
polymers in spin-coated films}.
\newblock \emph{\bibinfo{journal}{Phys. Rev. Lett.}}
\textbf{\bibinfo{volume}{99}}, \bibinfo{pages}{036101}  (\bibinfo{year}{2007}).

\bibitem {culick1960comments}\bibinfo{author}{Culick, F.}
\newblock \bibinfo{title}{Comments on a ruptured soap film}.
\newblock \emph{\bibinfo{journal}{J. Appl. Phys.}}
\textbf{\bibinfo{volume}{31}}, \bibinfo{pages}{1128--1129}  (\bibinfo{year}{1960}).

\bibitem {Mysels1969bursting}\bibinfo{author}{McEntee, W.~R.} \&
\bibinfo{author}{Mysels, K.~J.}
\newblock \bibinfo{title}{Bursting of soap films. i. an experimental study}.
\newblock \emph{\bibinfo{journal}{J. Phys. Chem.}}
\textbf{\bibinfo{volume}{73}}, \bibinfo{pages}{3018--3028}  (\bibinfo{year}{1969}).

\bibitem {frankel1969bursting}\bibinfo{author}{Frankel, S.} \&
\bibinfo{author}{Mysels, K.~J.}
\newblock \bibinfo{title}{Bursting of soap films. ii. theoretical
considerations}. \newblock \emph{\bibinfo{journal}{J. Phys. Chem.}}
\textbf{\bibinfo{volume}{73}}, \bibinfo{pages}{3028--3038}  (\bibinfo{year}{1969}).

\bibitem {EtienneBurstingInOil}\bibinfo{author}{Reyssat, E.} \&
\bibinfo{author}{Quere, D.}
\newblock \bibinfo{title}{Bursting of a fluid film in a viscous environment}.
\newblock \emph{\bibinfo{journal}{EPL (Europhys. Lett.)}}
\textbf{\bibinfo{volume}{76}}, \bibinfo{pages}{236} (\bibinfo{year}{2006}).

\bibitem {DewetFrancoisePRL91}\bibinfo{author}{Redon, C.},
\bibinfo{author}{Brochard-Wyart, F.} \&  \bibinfo{author}{Rondelez, F.}
\newblock \bibinfo{title}{Dynamics of dewetting}.
\newblock \emph{\bibinfo{journal}{Phys. Rev. Lett.}}
\textbf{\bibinfo{volume}{66}}, \bibinfo{pages}{715--718}  (\bibinfo{year}{1991}).

\bibitem {debregeas1995viscous}\bibinfo{author}{Debr{\'e}geas, G.},
\bibinfo{author}{Martin, P.} \&  \bibinfo{author}{Brochard-Wyart, F.}
\newblock \bibinfo{title}{Viscous bursting of suspended films}.
\newblock \emph{\bibinfo{journal}{Phys. Rev. Lett.}}
\textbf{\bibinfo{volume}{75}}, \bibinfo{pages}{3886} (\bibinfo{year}{1995}).

\bibitem {EriOkumura2010}\bibinfo{author}{Eri, A.} \&
\bibinfo{author}{Okumura, K.}
\newblock \bibinfo{title}{Bursting of a thin film in a confined geometry:
Rimless and constant-velocity dewetting}.
\newblock \emph{\bibinfo{journal}{Phys. Rev. E}} \textbf{\bibinfo{volume}{82}}%
,  \bibinfo{pages}{030601(R)} (\bibinfo{year}{2010}).

\bibitem {SquiresQuake2005}\bibinfo{author}{Squires, T.~M.} \&
\bibinfo{author}{Quake, S.~R.}
\newblock \bibinfo{title}{Microfluidics: Fluid physics at the nanoliter scale}.
\newblock \emph{\bibinfo{journal}{Rev. Mod. Phys.}}
\textbf{\bibinfo{volume}{77}}, \bibinfo{pages}{977} (\bibinfo{year}{2005}).

\bibitem {PhysicsFoams}\bibinfo{author}{Weaire, D.} \&
\bibinfo{author}{Hutzler, S.}
\newblock \emph{\bibinfo{title}{The Physics of Foams}}
(\bibinfo{publisher}{Clarendon Press, Oxford}, \bibinfo{year}{1999}).
\end{thebibliography}

\clearpage

\end{document}